\documentclass[aps,prapplied,showpacs,twocolumn,superscriptaddress,floatfix,preprintnumbers,longbibliography]{revtex4-2}
\usepackage{graphicx}
\usepackage{dcolumn}
\usepackage{bm}
\usepackage{braket}
\usepackage{hyperref}
\usepackage{amsmath}
\usepackage{amsthm,amssymb}
\usepackage{mathrsfs}
\usepackage{color}
\usepackage{mathtools}
\let\OLDthebibliography\thebibliography
\renewcommand\thebibliography[1]{
  \OLDthebibliography{#1}
  \setlength{\parskip}{0pt}
  \setlength{\itemsep}{0pt plus 0.1ex}
}
\hypersetup{colorlinks=true, linkcolor=blue, anchorcolor=blue, urlcolor=blue, citecolor=blue}
\renewcommand{\eqref}[1]{\hyperref[#1]{(}\ref{#1}\hyperref[#1]{)}}
\newcommand{\g}{^3A_2}
\newcommand{\e}{^3E}

\begin{document}

\title{Nonlinear Wave-Spin Interactions in Nitrogen-Vacancy Centers}

\author{Zhongqiang Hu}
\affiliation{Department of Electrical Engineering and Computer Science, Massachusetts Institute of Technology, Cambridge, MA 02139, USA}

\author{Qiuyuan Wang}
\affiliation{Department of Electrical Engineering and Computer Science, Massachusetts Institute of Technology, Cambridge, MA 02139, USA}

\author{Chung-Tao Chou}
\affiliation{Department of Electrical Engineering and Computer Science, Massachusetts Institute of Technology, Cambridge, MA 02139, USA}
\affiliation{Department of Physics, Massachusetts Institute of Technology, Cambridge, MA 02139, USA}

\author{Justin T. Hou}
\affiliation{Department of Electrical Engineering and Computer Science, Massachusetts Institute of Technology, Cambridge, MA 02139, USA}

\author{Zhiping He}
\affiliation{Department of Electrical Engineering and Computer Science, Massachusetts Institute of Technology, Cambridge, MA 02139, USA}

\author{Luqiao Liu}
\email{luqiao@mit.edu}
\affiliation{Department of Electrical Engineering and Computer Science, Massachusetts Institute of Technology, Cambridge, MA 02139, USA}

\date{\today}

\begin{abstract}

Nonlinear phenomena represent one of the central topics in the study of wave-matter interactions and constitute the key blocks for various applications in optical communication, computing, sensing, and imaging. In this work, we show that by employing the interactions between microwave photons and electron spins of nitrogen-vacancy (NV) centers, one can realize a variety of nonlinear effects, ranging from the resonance at the sum or difference frequency of two or more waves to electromagnetically induced transparency from the interference between spin transitions. We further verify the phase coherence through two-photon Rabi-oscillation measurements. The highly sensitive, optically detected NV-center dynamics not only provides a platform for studying magnetically induced nonlinearities but also promises novel functionalities in quantum control and quantum sensing.

\end{abstract}

\maketitle

\section{Introduction}

Through the mixing of multiple electromagnetic waves \cite{Shen1984, Boyd2008, Brabec2000, Reshef2019}, nonlinear processes provide useful mechanisms for frequency up- and down-conversion \cite{Fejer1994, Caspani2011, Guo2016, Li2016, Wang2021}, parametric signal amplification or generation \cite{Radic2003, Li2015, Reimer2015, Mosca2018}, as well as the creation of entangled photons or squeezed light \cite{Loudon1987, suhara2009, Quesada2015, Caspani2017}, the fundamental components of quantum information systems. For nonlinear interactions between waves and matter, electric dipole transitions are generally considered over their magnetic counterparts due to their larger strengths \cite{Buckholtz2020}. However, restrained by optical selection rules, special crystals with broken inversion symmetry are usually required for nonlinear coefficients such as the second-order electric susceptibility to be nonvanishing \cite{Shen1984, Boyd2008}. On the other hand, magnetic dipole transitions can possess nonlinearities even in centrosymmetric systems due to the inherent breaking of time-reversal symmetry. Nonlinear magnetic dipole transitions, particularly nonlinear spin transitions, have been touched upon in nuclear magnetic resonance (NMR) and electron paramagnetic resonance (EPR) \cite{Orton1960, Clerjaud1982, Zur1983, Kalin2006}, where more than one electromagnetic wave source is used for exciting the resonance. However, due to the very weak wave-spin interactions, these nonlinear signals are generally difficult to detect. To ensure measurable resonance, very low frequencies -- in the kilohertz or low-megahertz range -- have to be used for at least one of the input sources, making these measurements effectively the same as the field-modulation scheme of magnetic resonance. Therefore, a comprehensive study on multiphoton spin transitions that cover a broad frequency range and that can lead to useful quantum control and sensing protocols is highly desirable. 

The nitrogen-vacancy (NV) center, an extensively studied quantum defect in diamond, has been pursued as a magnetometer with fine spatial resolution and high sensitivity \cite{Taylor2008, Balasubramanian2008, Acosta2009, Maertz2010, Wolf2015, Barry2020}, and as a qubit for quantum information processing \cite{Dutt2007, Neumann2010, Togan2010, Fuchs2011, Childress2013, Bernien2013, Humphreys2018}. To achieve quantum state control, existing studies focus on linear processes by applying gigahertz microwaves at or close to the intrinsic resonance frequency. Recently, quantum frequency mixing based on sophisticated Floquet Hamiltonian engineering has been developed for magnetic field sensing with NV centers \cite{Wang2022}. The magnetic field at 150 MHz has been detected using the difference frequency of two waves through a spin-locked sensing protocol, under the assistance of a third, control signal at the original resonance frequency. In NV-center resonance, the detection of photons in the visible-light region rather than those in the radio-frequency or microwave domains greatly enhances the sensitivity, and leads to a superior platform for studying nonlinear spin transitions. In this work, we demonstrate such opportunities by carrying out a systematic study on nonlinear wave-spin interactions in NV centers. We show that the nonlinear resonance condition can be reached over a broad frequency range, at the sum or difference frequency of two waves, as well as with higher-order effects involving three, four, or more photons. Utilizing the interference between spin transitions, we further show that the resonance can be greatly suppressed in the presence of a probe wave and a strong control wave, leading to electromagnetically induced transparency (EIT). Finally, on top of continuous-wave measurements, we also observe sum-frequency Rabi oscillations, which not only verifies the phase coherence of these multiphoton processes but also suggests new mechanisms for quantum control and sensing.

\section{Optical detection of multiphoton spin transitions}

\begin{figure}[t!]
\centering	\includegraphics[width=1\columnwidth]{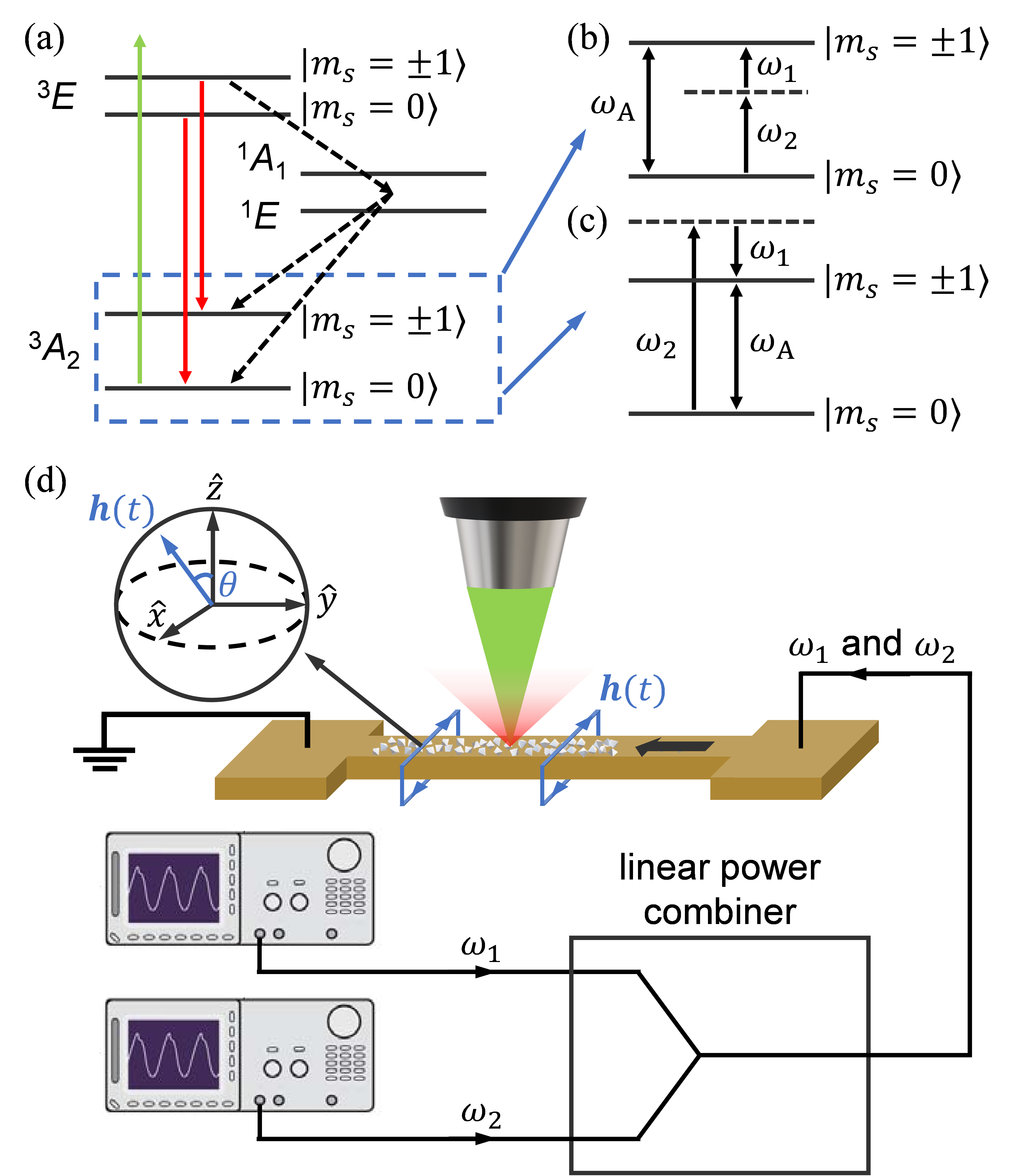}
\caption{(a) The energy-level diagram of an NV center. (b)–(c) Nonlinear spin transitions in the ground state $\g$ when the (b) sum or (c) difference of two applied microwave frequencies $\omega_1$ and $\omega_2$ matches the transition frequency $\omega_\mathrm{A}$. (d) A schematic of nonlinear ODMR measurements. Microwaves from two signal generators are combined through a linear power combiner, and applied onto a copper microstrip, exciting NV-center resonance in microdiamonds on top of the strip. The top-left inset shows an illustration of the geometrical relationship between microwave magnetic fields $\bm{h}(t)$ and an NV spin, where an angle of $\theta$ is formed between $\bm{h}(t)$ and the principal spin axis ($z$ axis).}
\label{fig1}
\end{figure}

Figure \hyperref[fig1]{1(a)} illustrates the energy levels of an NV center, where both the optical ground state $\g$ and excited state $\e$ are spin triplets with spin sublevels of $\ket{m_s = 0}$ and $\ket{m_s = \pm1}$ separated by $\omega_\mathrm{A} / (2\pi) = 2.87 \ \mathrm{GHz}$ and $\omega_\mathrm{E} / (2\pi) = 1.42 \ \mathrm{GHz}$, respectively, under zero external static field \cite{Doherty2012, Rogers2009}. Green light can induce the transition from $\g$ to $\e$, and the $m_s$-conserving decay from $\e$ to $\g$ generates photoluminescence (PL) in the region of red light \cite{Jensen2013}. The nonradiative transition path through spin singlet states $^1A_1$ and $^1E$ pumps the NV population into the $\ket{m_s = 0}$ sublevel, which can be suppressed with the application of a microwave at or close to the sublevel splitting $\omega_\mathrm{A}$ or $\omega_\mathrm{E}$, yielding a reduction of the PL intensity \cite{Rogers2008, Acosta2010, Choi2012}. In this work, we will delve into optically detected magnetic resonance (ODMR) beyond the linear response regime and investigate spin transitions induced by multiple photons, through concurrent application of two or more microwaves. In Figs. \hyperref[fig1]{1(b)} and \hyperref[fig1]{1(c)}, we illustrate two example scenarios where the sum or difference of the two applied frequencies $\omega_1$ and $\omega_2$ matches the ground-state transition frequency $\omega_\mathrm{A}$. To excite magnetic resonance in experiments, microwaves from two independent signal generators are combined through a power combiner [Fig. \hyperref[fig1]{1(d)}]. We have verified that under our employed power levels, the external microwave circuit acts purely linearly and is not the origin of frequency mixing (see Appendix \ref{appA}). The microwaves are further applied onto a lithographically defined copper microstrip on a silicon substrate. Diamond particles with a diameter of approximately $1 \ \rm \mu m$ and an NV-center concentration of approximately $3.5 \ \rm ppm$ (parts per million) are dispersed on top of the strip. PL excited by a $532$-$\rm nm$ green laser is filtered and collected with a photomultiplier tube. 

\begin{figure}[t]
\centering	\includegraphics[width=1\columnwidth]{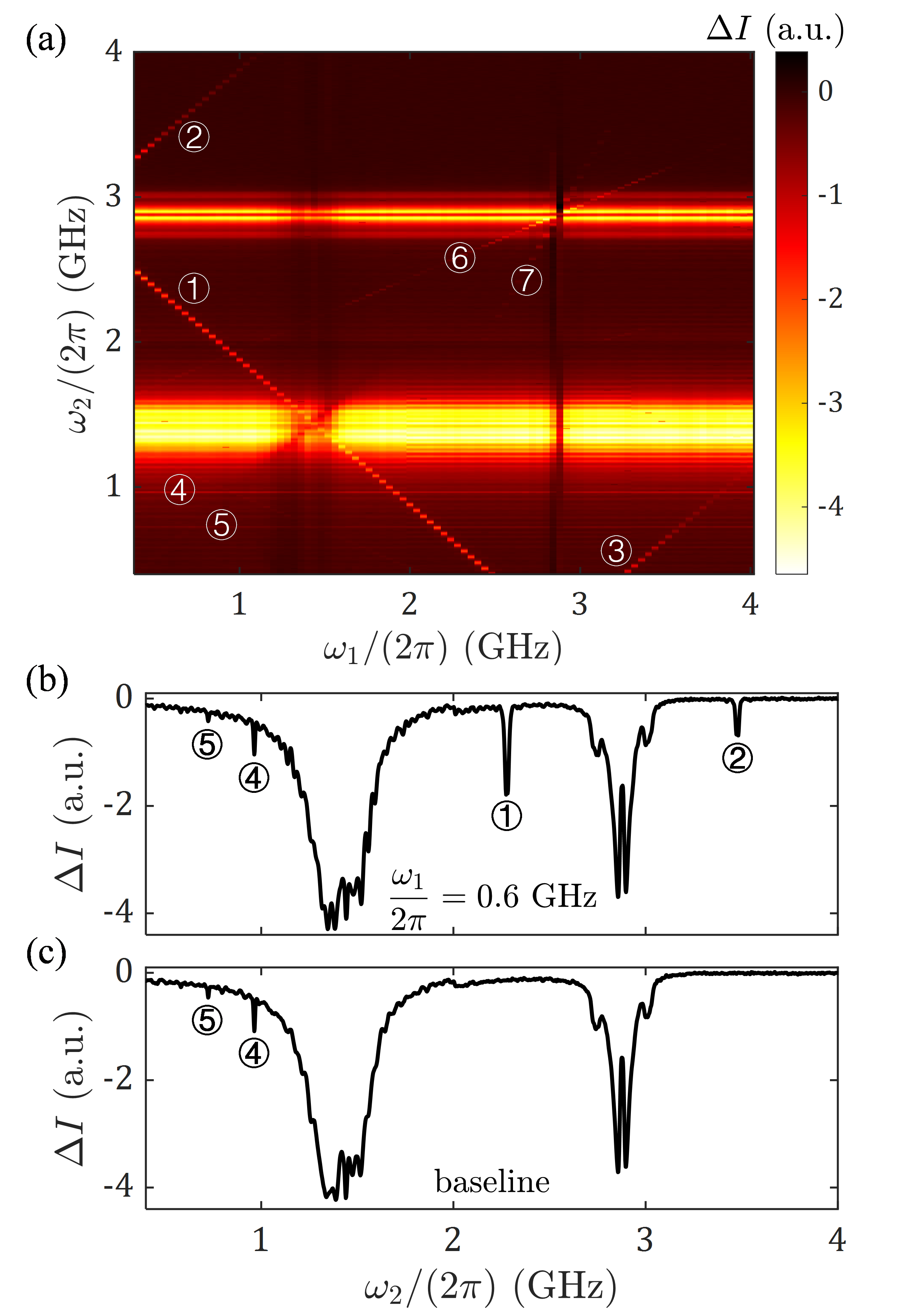}
\caption{(a) The change of the PL intensity $\Delta I$ under driving microwaves of $\omega_1$ and $\omega_2$ with input powers $P_1 = P_2 = 13 \ \mathrm{mW}$. Nonlinear resonance signals emerge at (1) $\omega_2 + \omega_1 = \omega_\mathrm{A}$, (2) $\omega_2 - \omega_1 = \omega_\mathrm{A}$, (3) $\omega_1 - \omega_2 = \omega_\mathrm{A}$, (4) $3\omega_2 = \omega_\mathrm{A}$, (5) $4\omega_2 = \omega_\mathrm{A}$, (6) $2\omega_2 - \omega_1 = \omega_\mathrm{A}$, and (7) $2\omega_1 - \omega_2 = \omega_\mathrm{A}$. (b) $\omega_2$ scan in (a) when $\omega_1 / (2\pi) = 0.6 \ \mathrm{GHz}$. (c) The $\omega_2$ scan when $P_1 = 0$, serving as the baseline of the measurement. The splitting of the linear resonance dip at $\omega_2 = \omega_\mathrm{A}$ (red arrow) is a result of amplitude modulation at high applied microwave powers, which disappears in unmodulated results (see Appendix \ref{appA}). The side dips denoted by green arrows originate from the interactions between NV centers and P1 centers \cite{Kamp2018}.}
\label{fig2}
\end{figure}

In Fig. \hyperref[fig2]{2(a)}, we show the change of the PL intensity $\Delta I$ under driving microwaves of $\omega_1$ and $\omega_2$. To enhance the signal-to-noise ratio, we modulate the amplitude of the $\omega_2$ input with a frequency of $104.42 \ \mathrm{Hz}$ and detect $\Delta I$ using a lock-in amplifier. We have compared results from this lock-in measurement with a standard unmodulated continuous-wave measurement (see Appendix \ref{appA}) and we have confirmed that these two give the consistent results and that the low-frequency amplitude modulation is not the source of the observed nonlinear effects. Since the amplitude modulation only acts on the $\omega_2$ input, it gives rise to the asymmetry on the dependence of $\Delta I$ with respect to $\omega_{1, 2}$ in Fig. \hyperref[fig2]{2(a)}: resonance signals only show up at $\omega_2 = \omega_\mathrm{A, E}$ but not at $\omega_1 = \omega_\mathrm{A, E}$, in contrast to the unmodulated measurement results. The large line widths associated with the $\omega_2 = \omega_\mathrm{A,E}$ resonance dips reflect the applied high microwave power and low laser pump power (about 0.6 mW) \cite{Jensen2013, Dreau2011}. Besides the standard, linear resonance dips, in Fig. \hyperref[fig2]{2(a)} additional resonance signals emerge when $\omega_{1,2}$ satisfy the relationship of $\omega_2 + \omega_1 = \omega_\mathrm{A}$ (labeled as 1) or $\pm(\omega_2 - \omega_1) = \omega_\mathrm{A}$ (labeled as 2 and 3). As an example, we show the spectrum when $\omega_1/2\pi$ is fixed at $0.6 \ \mathrm{GHz}$ and $\omega_2$ is swept [Fig. \hyperref[fig2]{2(b)}] and compare it with the baseline when the power of the $\omega_1$ input is set to zero [Fig. \hyperref[fig2]{2(c)}]. In Fig. \hyperref[fig2]{2(b)}, the depths of the $\omega_2 + \omega_1 = \omega_\mathrm{A}$ and $\omega_2 - \omega_1 = \omega_\mathrm{A}$ dips reach around 50\% and 20\% of that of the $\omega_2 = \omega_\mathrm{A}$ dip. The two nonlinear resonance dips can be easily detected even when $\omega_{1, 2}$ are individually far away from $\omega_\mathrm{A}$.

\section{Theoretical model for multiphoton spin transitions}

To understand the origin of magnetic resonance occurring at the sum or difference frequency, we next model the NV spin transitions driven by multiple microwave photons. Here, we consider, e.g., the spin transition between $\ket{m_s = 0}$ and $\ket{m_s = +1}$ in $\g$. Other transitions, such as that between $\ket{m_s = 0}$ and $\ket{m_s = -1}$ and those in $\e$, can be treated similarly. We write the Hamiltonian of the spin-photon system as
\begin{equation}
    H(t) = \frac{\hbar \omega_\mathrm{A}}{2} \sigma_z + \gamma \mu_0 \frac{\hbar}{2} \bm{\sigma} \cdot \bm{h}(t),
\label{Hamiltonian}
\end{equation}

\noindent where $\hbar$ is reduced Planck constant, $\bm{\sigma} = (\sigma_x, \sigma_y, \sigma_z)$ are Pauli matrices, $\gamma$ is the electron's gyromagnetic ratio, $\mu_0$ is the vacuum permeability, and $\smash{\bm{h}(t) = \sum_{j=1,2}\bm{h}_j(t)}$ are microwave fields with the $j$th frequency component $\bm{h}_j(t)$. In experiments, the two microwave fields are launched by the same microstrip; thus $\bm{h}_1(t)$ and $\bm{h}_2(t)$ are collinear and form an angle $\theta$ with the principal axis of a given NV spin [see the top-left inset of Fig. \hyperref[fig1]{1(d)}], $\bm{h}_j(t) = h_j (\sin\theta \hat{x} + \cos\theta \hat{z}) \cos(\omega_j t + \varphi_j)$, where $h_j$, $\omega_j$, and $\varphi_j$ are the amplitude, frequency, and phase of the $j$th field ($j = 1, 2$).

Solving the quantum master equations iteratively (see the derivation details in Appendixes \ref{appB} and \ref{appC}), we obtain the change of the PL intensity for an ensemble of spins when the sum or difference of $\omega_{1,2}$ is close to the resonance condition of $\omega_2 \pm \omega_1 = \omega_\mathrm{A}$:
\begin{equation}
    \Delta I = - \eta |\chi_{xxz}^{(2)}(\omega_2 \pm \omega_1, \omega_1, \omega_2)|^2 h_1^2 h_2^2.
\label{Delta I main}
\end{equation}

\noindent Here, $\eta = |\Delta I|_\mathrm{max} \Gamma_\mathrm{2, A} / (16 \gamma^2 \hbar^2 \Gamma_\mathrm{p})$ is a factor depending on the maximum PL intensity $|\Delta I|_\mathrm{max}$, the effective transverse relaxation rate $\Gamma_\mathrm{2, A}$ in $\g$, as well as the laser pump rate $\Gamma_\mathrm{p}$. $\smash{\chi_{xxz}^{(2)}}$, defined through $M_x = \smash{\chi_{xxz}^{(2)}} h_x h_z$, is an element in the second-order magnetic susceptibility tensor for a single spin, where $M_x$ is the $x$-axis component of the magnetic moment and $\smash{h_{x(z)}}$ is the $x(z)$-axis component of $\bm{h}(t)$. Close to the resonance, we have
\begin{equation}
    \chi_{xxz}^{(2)}(\omega_2 \pm \omega_1, \omega_1, \omega_2) = \mp \frac{\Gamma_\mathrm{p} \gamma^3 \mu_0^2 \hbar \omega_\mathrm{A}}{4\Gamma_\mathrm{1, A} (\Delta_\mathrm{\pm, A} - i \Gamma_\mathrm{2, A}) \omega_1 \omega_2}, 
\label{Chi}
\end{equation}

\noindent where the $+ \ (-)$ sign is chosen when the resonance condition is satisfied by the sum(difference) frequency, $\Gamma_\mathrm{1, A}$ is the effective longitudinal relaxation rate in $\g$, and $\Delta_\mathrm{\pm, A} = \omega_2 \pm \omega_1 - \omega_\mathrm{A}$ denotes the frequency detuning. We can infer important information on the two-photon spin transition from $\smash{\chi_{xxz}^{(2)}}$. First, $\smash{\chi_{xxz}^{(2)}}$ remains finite irrespective of the atom's inversion symmetry, in contrast to the second-order electric susceptibility which vanishes for atoms with inversion symmetry \cite{Shen1984, Boyd2008}. Second, the fact that the magnetic moment resonates at the sum or difference frequency of the input sources reflects the energy conservation. Third, angular-momentum conservation is also respected in the two-photon spin transition described by $\smash{\chi_{xxz}^{(2)}}$: the perpendicular microwave field $h_x$ comprises circularly polarized photons $\sigma^{\pm}$, each carrying an angular momentum of $\pm \hbar$, while the parallel component $h_z$ comprises $\pi$ photons that carry zero angular momentum \cite{Kalin2006}. Therefore, the transition from $\ket{m_s = 0}$ to $\ket{m_s = \pm1}$ via $\sigma^{\pm} + \pi$ leaves the total angular momentum of the spin-photon system unchanged.

\begin{figure*}[t]
\centering	\includegraphics[width=1\textwidth]{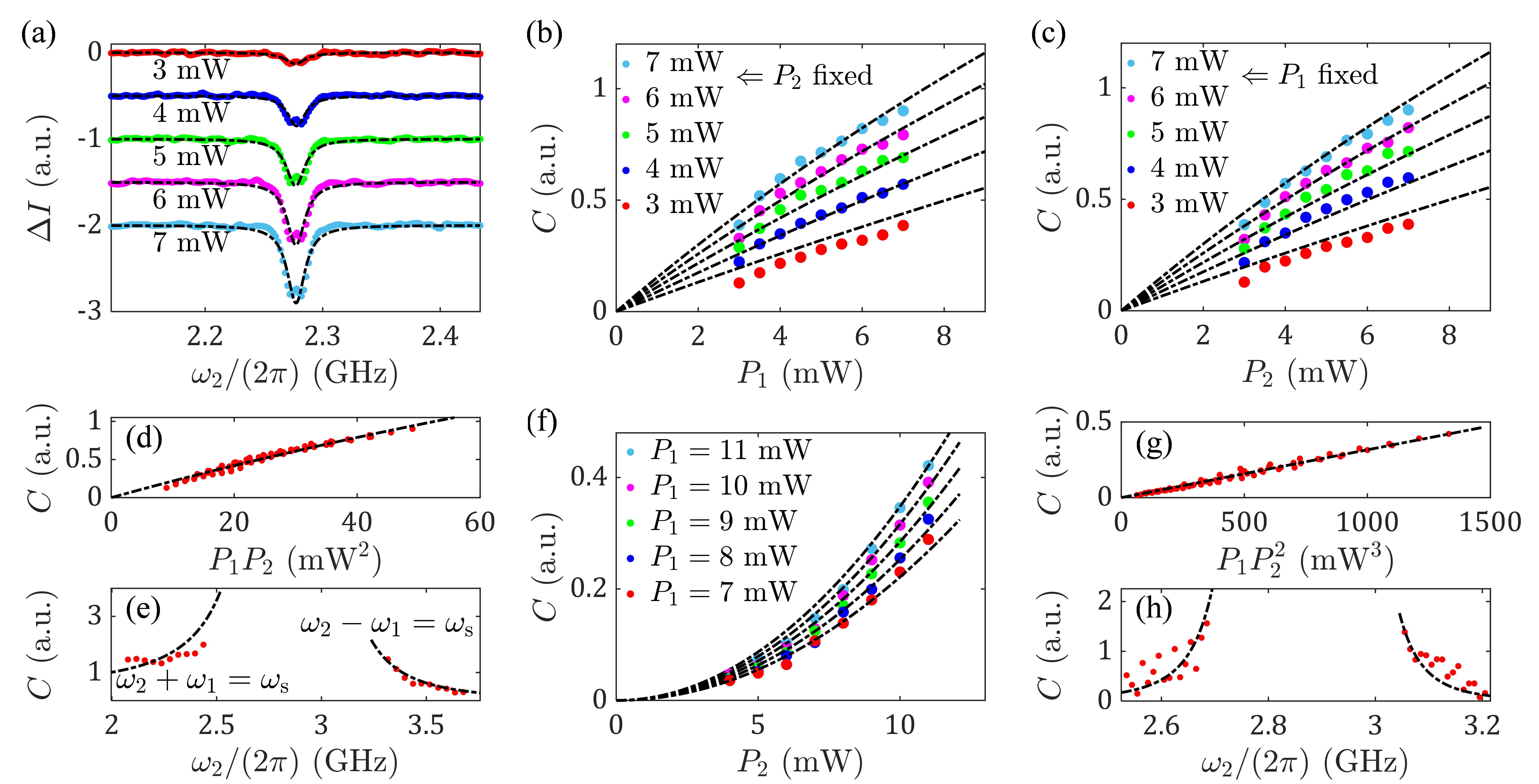}
\caption{(a) The sum-frequency resonance under different input microwave powers ($P_1 = P_2$ is maintained), when $\omega_1/(2\pi) = 0.6 \ \rm GHz$. Neighboring curves and data are shifted by $-0.5$ for clear visualization. (b) The contrast $C$ of the sum-frequency resonance as a function of $P_1$ when $P_2$ is fixed. (c) The $C$ value of the sum-frequency resonance as a function of $P_2$ when $P_1$ is fixed. (d) The $C$ value of the sum-frequency resonance as a function of $P_1 P_2$. (e) The $C$ value of the sum- or difference-frequency resonance as a function of $\omega_2$ ($P_1 = P_2 = 9 \ \rm mW$). Dashed lines in (b)–(e) represent fitting with Eq. \eqref{peak height main} using the same parameter $\eta$. (f) The $C$ value of the three-photon resonance at $2\omega_2 - \omega_1 = \omega_\mathrm{A}$ as a function of $P_2$ when $P_1$ is fixed. (g) The $C$ value of the $2\omega_2 - \omega_1 = \omega_\mathrm{A}$ resonance as a function of $P_1 P_2^2$. (h) The $C$ value of the $2\omega_2 - \omega_1 = \omega_\mathrm{A}$ resonance as a function of $\omega_2$ ($P_1 = P_2 = 11 \ \rm mW$). The dashed lines in (f)-(h) represent fitting with $\smash{\chi^{(3)}}$, the expression for which is derived in Appendix \ref{appD}.}
\label{fig3}
\end{figure*}

We next compare our measurement results quantitatively with the proposed theory. Equation \eqref{Delta I main} represents a Lorentzian with a line width of $\Gamma_\mathrm{2, A}$ and a contrast of 
\begin{equation}
    C = \frac{|\Delta I|_\mathrm{max} \Gamma_\mathrm{p} \omega_\mathrm{A}^2 \gamma^4 \mu_0^4 h_1^2 h_2^2}{256 \Gamma_\mathrm{1, A}^2 \Gamma_\mathrm{2, A} \omega_1^2 \omega_2^2}.
\label{peak height main}
\end{equation}

\noindent In Fig. \hyperref[fig3]{3(a)}, we show a series of resonance curves under varying input microwave powers $P_{1,2}$, when $\omega_1/(2\pi)$ is fixed at $ 0.6 \ \rm GHz$ and $\omega_2$ is swept around $\omega_\mathrm{A} - \omega_1$. The line shape in Eq. \eqref{Delta I main} (dashed lines) agrees well with the experimental results (solid circles). The resonance contrast $C$ obtained under different combinations of $P_{1,2}$ is summarized in Figs. \hyperref[fig3]{3(b)–3(d)}, in which the theoretical curves from Eq. \eqref{peak height main} (dashed lines) confirm the linear relationship between $C$ and $P_1 P_2$. The same value of $\eta$, the only fitting parameter, is used across all the curves. The influence from frequencies of the two applied microwaves is presented in Fig. \hyperref[fig3]{3(e)}, where $\omega_{1,2}$ are varied simultaneously, and their sum or difference is maintained at $\omega_2 \pm \omega_1 = \omega_\mathrm{A}$. $C$ has a frequency dependence of $\omega_2^{-2}(\omega_\mathrm{A} - \omega_2)^{-2}$, consistent with Eq. \eqref{peak height main}. The fitting curves (dashed lines) use the same value of $\eta$ as in Figs. \hyperref[fig3]{3(b)–3(d)}.

Besides the two-photon resonance investigated above, in Fig. \hyperref[fig2]{2(a)} we observe additional bright lines (labeled from 4 to 7), which can be traced to magnetic resonance excited by even higher-order processes. Our examination shows that signals of 4, 6, and 7 satisfy the frequency relationship of (4) $3\omega_2 = \omega_\mathrm{A}$, (6) $2\omega_2 - \omega_1 = \omega_\mathrm{A}$, and (7) $2\omega_1 - \omega_2 = \omega_\mathrm{A}$, corresponding to three-photon processes. The horizontal line with a resonance frequency of $0.72 \ \mathrm{GHz}$ (labeled as 5) corresponds to four-photon resonance with $4\omega_2 = \omega_\mathrm{A}$. These higher-order processes can be well described with the theoretical framework that we have developed, by calculating higher-order magnetic susceptibilities. For example, in Figs. \hyperref[fig3]{3(f)} and \hyperref[fig3]{3(g)} we summarize the dependence on the input powers for the three-photon resonance at $2\omega_2 - \omega_1 = \omega_\mathrm{A}$. As indicated by $\smash{\chi^{(3)}}$ (see Appendix \ref{appD}), the contrast of this resonance has a quadratic dependence on $P_2$ and a linear dependence on $P_1$. The influence from frequencies of the two applied microwaves is presented in Fig. \hyperref[fig3]{3(h)}, which also fits well with the theory. Finally, we note that the $2\omega_2 = \omega_\mathrm{A}$ line, as a special case of the sum-frequency resonance, falls onto the broad $\omega_2 = \omega_\mathrm{E}$ resonance dip and is difficult to distinguish due to the small difference between $\omega_\mathrm{A} / 2$ and $\omega_\mathrm{E}$.

While the results in Figs. \hyperref[fig2]{2} and \hyperref[fig3]{3} correspond to the scenario with zero external static field, we have verified that the nonlinear processes remain in the presence of a finite static field that lifts the degeneracy in $\ket{m_s = \pm1}$ (see Appendix \ref{appA}).

\section{Electromagnetically induced transparency in NV centers}

\begin{figure*}[t!]
\centering	\includegraphics[width=1\textwidth]{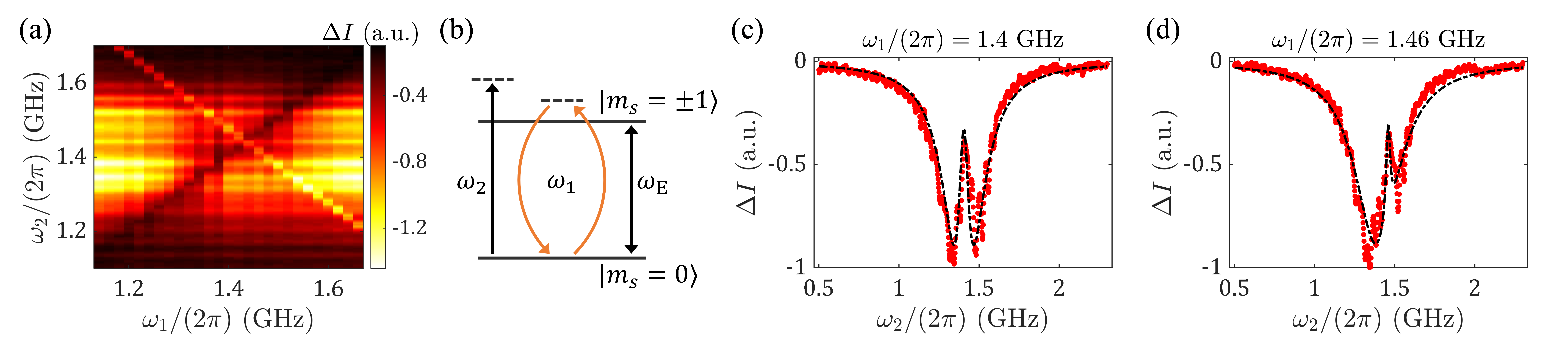}
\caption{A demonstration of the EIT effect. (a) $\Delta I$ under $\omega_1$ and $\omega_2$ inputs when $\omega_{1,2}$ are close to $\omega_\mathrm{E}$. The microwave powers are $P_1 = 13 \ \mathrm{mW}$, $P_2 = 5 \ \mathrm{mW}$. (b) Destructive interference between the spin transition with $\omega_2$ absorption [associated with $\smash{\chi^{(1)}}$; see Appendix \ref{appE}] and that with $\omega_2$ absorption, $\omega_1$ emission, $\omega_1$ absorption [associated with $\smash{\chi^{(3)}}$]. (c)–(d) $\Delta I$ as a function of $\omega_2$ when $\omega_1 / (2\pi)$ is fixed at (c) $1.4 \ \rm GHz$ and (d) $1.46 \ \rm GHz$. The dashed lines represent fitting with Eq. \eqref{EIT_main}.}
\label{fig4}
\end{figure*}

In Fig. \hyperref[fig2]{2(a)}, on top of the series of the bright resonance lines, we observe a dark line satisfying $\omega_1 = \omega_2$ within the broad resonance dip at $\omega_2 = \omega_\mathrm{E}$. This feature of magnetic resonance suppression under zero detuning of two waves is very similar to the electromagnetically induced transparency (EIT) phenomenon studied in nonlinear optics \cite{Fleischhauer2005}, where in the presence of a strong pump wave, the interaction between the probe wave and the matter is minimized due to interference effects. This signal-suppression phenomenon has also been reported before under the concept of coherent population oscillation or trapping, for nuclear and electron spin systems \cite{Jamonneau2016, Mrozek2016}. Treating the $\omega_1$ signal as the pump and the modulated $\omega_2$ signal as the probe, we examine the resonance results with a weaker probe power $P_2$ while maintaining a high pump power $P_1$ and find that the transparency window still exists [Fig. \hyperref[fig4]{4(a)}]. Examples of $\omega_2$ scans are presented in Figs. \hyperref[fig4]{4(c)} and \hyperref[fig4]{4(d)}. On the other hand, the transparency window disappears when a low pump power $P_1$ is used (see Appendix \ref{appE}). The magnetic resonance suppression under $\omega_1 = \omega_2$ can be explained by considering the destructive interference between the $\ket{m_s = 0} \xrightarrow{\omega_2} \ket{m_s = \pm1}$ spin transition associated with a first-order susceptibility and the $\ket{m_s = 0} \xrightarrow{\omega_2} \ket{m_s = \pm1} \smash{\xrightarrow{-\omega_1}} \ket{m_s = 0} \xrightarrow{\omega_1} \ket{m_s = \pm1}$ spin transition associated with a third-order susceptibility [Fig. \hyperref[fig4]{4(b)}]. Mathematically, when $\omega_1 \approx \omega_2$ and both of them are close to $ \omega_\mathrm{E}$, we have (see the derivation details in Appendix \ref{appE})
\begin{equation}
    \Delta I =
    \frac{iA}{\omega_2 - \omega_\mathrm{E} - i\Gamma_\mathrm{2, E} - \frac{B}{\omega_2 - \omega_1 - i\Gamma_\mathrm{1, E}}} + \mathrm{H.c.},
\label{EIT_main}
\end{equation}

\noindent where $A = |\Delta I|_\mathrm{max} \Gamma_\mathrm{p} \gamma^2 \mu_0^2 h_2^2 / (32\Gamma_\mathrm{1, E}^2)$, $B = \gamma^2 \mu_0^2 h_1^2 / 4$, and $\Gamma_\mathrm{1(2), E}$ is the effective longitudinal(transverse) relaxation rate in $\e$. In Eq. \eqref{EIT_main}, we see that the peak value of $|\Delta I|$ is suppressed at zero detuning $\omega_2 - \omega_1 = 0$ and that the EIT effect is most significant when $\Gamma_\mathrm{2, E} > \gamma \mu_0 h_1 / 2 > \Gamma_\mathrm{1, E}$, which is satisfied in the spin transitions in $\e$. Comparatively, the relatively smaller $\Gamma_\mathrm{2,A}$ makes the EIT feature in $\g$ less noticeable.

\section{Coherent control of NV spin state through multiple microwave photons}

\begin{figure*}[t]
\centering	\includegraphics[width=1.0\textwidth]{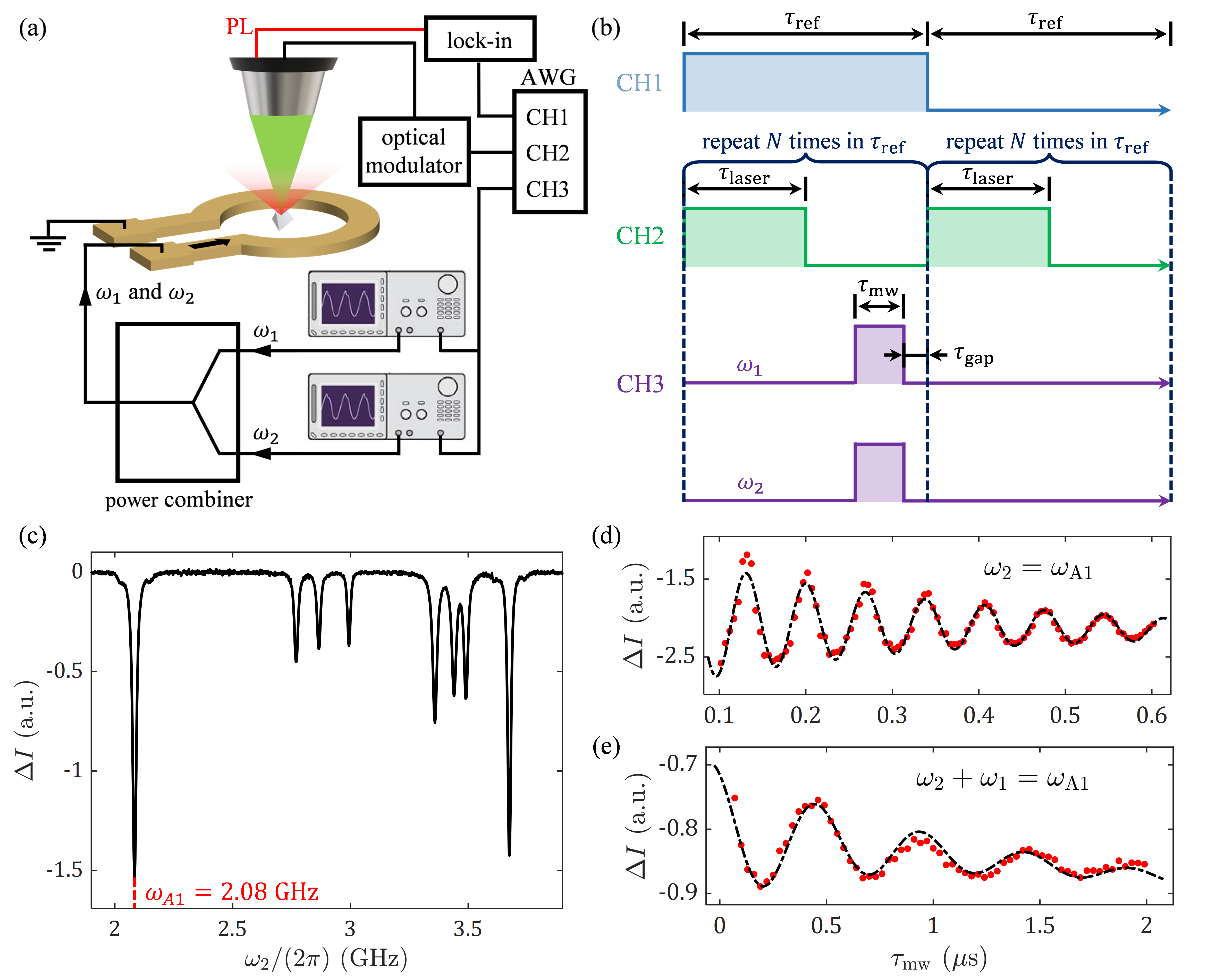}
\caption{(a) Experimental schematic for the two-photon Rabi-oscillation measurements. A single-crystal diamond with a diameter of 15 $\mathrm{\mu m}$ is placed at the center of a copper microstrip ring. The inner and outer diameter of the ring is 60 $\mathrm{\mu m}$ and 100 $\mathrm{\mu m}$, respectively. CH2 and CH3 of the AWG modulate the laser and two microwave sources, respectively, where CH1 provides a low-frequency (200 Hz) reference signal for the lock-in amplifier. (b) Pulse sequences for lock-in amplifier (blue), laser (green), and two microwave sources (violet). (c) ODMR spectrum in the presence of an external static field of 300 Oe under $\omega_2$ microwave source only, with $P_2 = 13 \ \mathrm{mW}$. The leftmost resonance at $\omega_\mathrm{A1} / (2\pi) = 2.08 \ \mathrm{GHz}$ is selected for Rabi-oscillation measurements. (d) Standard Rabi oscillation results under a single source $\omega_2 = \omega_\mathrm{A1}$ with $P_2 = 200 \ \mathrm{mW}$. The single-photon Rabi frequency is 14.5 MHz. (e) Rabi oscillation results when simultaneously applying $\omega_1 / (2\pi) = 60 \ \mathrm{MHz}$, $P_1 = 13 \ \mathrm{mW}$ and $\omega_2 = \omega_\mathrm{A1} - \omega_1$, $P_2 = 200 \ \mathrm{mW}$. The two-photon Rabi frequency is 2.03 MHz. In (d) and (e), the experimental data (solid circles) are fitted with Eq. \eqref{Rabi} (dashed lines).}
\label{fig5}
\end{figure*}

Up to now, the continuous-wave measurements as described above have allowed us to capture magnetic resonance signals at different orders for a broad frequency range. In the following, we carry out Rabi oscillation experiments to evaluate the phase coherence of the nonlinear multiphoton spin transitions, which is of paramount significance in developing effective quantum control and sensing protocols. The experimental schematic is shown in Fig. \hyperref[fig5]{5(a)}. A single-crystal diamond with a diameter of 15 $\mathrm{\mu m}$ is placed at the center of a copper microstrip ring. The inner and outer diameters of the ring are 60 $\mathrm{\mu m}$ and 100 $\mathrm{\mu m}$, respectively, such that the generated microwave fields are nearly constant within the focal spot of the objective lens, the diameter of which is $< 1 \ \mathrm{\mu m}$. The switch to an individual diamond and a ring-shaped microstrip is to minimize the inhomogeneity in the detected NV centers. An arbitrary waveform generator (AWG) is used to program the pulse sequences for the measurements. CH2 and CH3 modulate the laser and two microwave sources, respectively, where CH1 provides a low-frequency (200 Hz) reference signal for a lock-in amplifier. Figure \hyperref[fig5]{5(b)} shows the pulse sequences for the lock-in amplifier (blue), the laser (green), and two microwave sources (violet). In our experiments, $\tau_\mathrm{laser}$ is set at $5 \ \mathrm{\mu s}$, $\tau_\mathrm{mw}$ is varied between $0.1 \ \mathrm{\mu s}$ and $2 \ \mathrm{\mu s}$, and $\tau_\mathrm{gap}$ is fixed at $1 \ \mathrm{\mu s}$. The lock-in reference signal (CH1) is fixed at 200 Hz with $\tau_\mathrm{ref} = 2.5 \ \mathrm{ms}$. Within each ON half period of CH1, the laser and microwave pulses are repeated $N = 250$ times. Within each OFF half period of CH1, only the laser pulse is repeated with the microwave pulse always OFF. This lock-in-based pulse method avoids the necessity of high-frequency electronics for data acquisition and possesses high sensitivity \cite{Sewani2020}.

In Fig. \hyperref[fig5]{5(c)}, we show the single-frequency ODMR spectrum under an external static field of $H_\mathrm{ext} = 300 \ \mathrm{Oe}$. We see eight resonance dips, corresponding to four different NV-axis orientations and two spin transitions $\ket{m_s = 0} \rightarrow \ket{m_s = \pm 1}$ in the NV ground state $\g$. The leftmost dip, with a resonance frequency of $\omega_\mathrm{A1} / (2\pi) = 2.08 \ \mathrm{GHz}$, is selected for oscillation measurements. By applying a single microwave source $\omega_2$ at the exact resonance frequency and varying its pulse width, we summarize the oscillation results in Fig. \hyperref[fig5]{5(d)}, which are fitted with

\begin{equation}
    \Delta I = A \sin\left(2\pi f_\mathrm{R} \tau_\mathrm{mw} + \phi\right) \exp\left(-\frac{\tau_\mathrm{mw}}{T_2^*}\right) + B\tau_\mathrm{mw} + C,
\label{Rabi}
\end{equation}

\noindent where $A$ is the Rabi oscillation amplitude, $B$ is a linear coefficient that includes the heating caused shift during measurement, $C$ is the readout at the steady state, $f_\mathrm{R}$ is the Rabi frequency, $\phi$ is the phase offset, and $T_2^*$ is the transverse relaxation time. The single-photon Rabi frequency $\smash{f_\mathrm{R}^{(1)}}$ is determined to be $14.5 \ \mathrm{MHz}$. We next realize two-photon Rabi oscillations by applying both microwave sources $\omega_1 / (2\pi) = 60 \ \mathrm{MHz}$ and $\omega_2 / (2\pi) = 2.02 \ \mathrm{GHz}$, with synchronous pulse modulation. We have verified that the $\omega_1$ or $\omega_2$ source at these frequencies alone does not induce detectable oscillation signals. However, as they satisfy the sum-frequency resonance condition, their concurrent application leads to clear oscillations, as shown in Fig. \hyperref[fig5]{5(e)}, with a Rabi frequency of $\smash{f_\mathrm{R}^{(2)}} = 2.03 \ \mathrm{MHz}$. The phase coherence demonstrated by our Rabi-oscillation measurements hopefully paves the way for the use of multiphoton spin transitions in practical quantum control and sensing schemes.

\section{Conclusions}

In conclusion, we have conducted a systematic study on the nonlinear interactions between electromagnetic waves and electron spins of NV centers. With the high sensitivity of NV-center resonance, we have revealed various nonlinear, parametric processes over a broad frequency range, ranging from high-order magnetic resonance to EIT. We have developed a theoretical framework based on perturbation theory to account for these nonlinear phenomena quantitatively. In addition, we have verified the phase coherence of the multiphoton spin transitions through Rabi-oscillation measurements, which hopefully paves the way toward future applications in quantum control and sensing. Furthermore, leveraging the sum- or difference-frequency resonance, one can make a nanoscale spectrometer out of NV centers to extract the spectrum information of oscillating magnetic fields such as those from spin-wave excitations in magnetic materials \cite{Du2017, Fukami2021, Koerner2022}, extending their well-established role as a magnetometer.

\begin{acknowledgements}
The work is supported by the National Science Foundation under Award No. DMR-2104912. L.L. acknowledges support from the Sloan Research Fellowship. Z.H. acknowledges support from the MathWorks Fellowship.

The manuscript has been accepted by \href{https://journals.aps.org/prapplied/accepted/af07aY5aSbc15790246255933943cc983e2521e93}{Physical Review Applied} (APS copyright) on Mar 29, 2024.
\end{acknowledgements}

\appendix

\section{Experimental method}\label{appA}
\textit{Diamond samples}: The samples used for continuous-wave measurements are high-pressure and high-temperature (HPHT) microdiamonds with a diameter of 1 $\mathrm{\mu m}$ and an NV concentration of 3.5 ppm (MDNV1umHi, Adámas Nanotechnologies). The samples used for the Rabi-oscillation measurements are 15-$\mathrm{\mu m}$ diamond particles from the same company, with the same NV concentration and growth method (MDNV15umHi). The continuous-wave measurements have been carried out on a cluster of diamond particles, while the Rabi oscillations have been done on an individual diamond. 

\begin{figure}[b!]
\centering	\includegraphics[width=1\columnwidth]{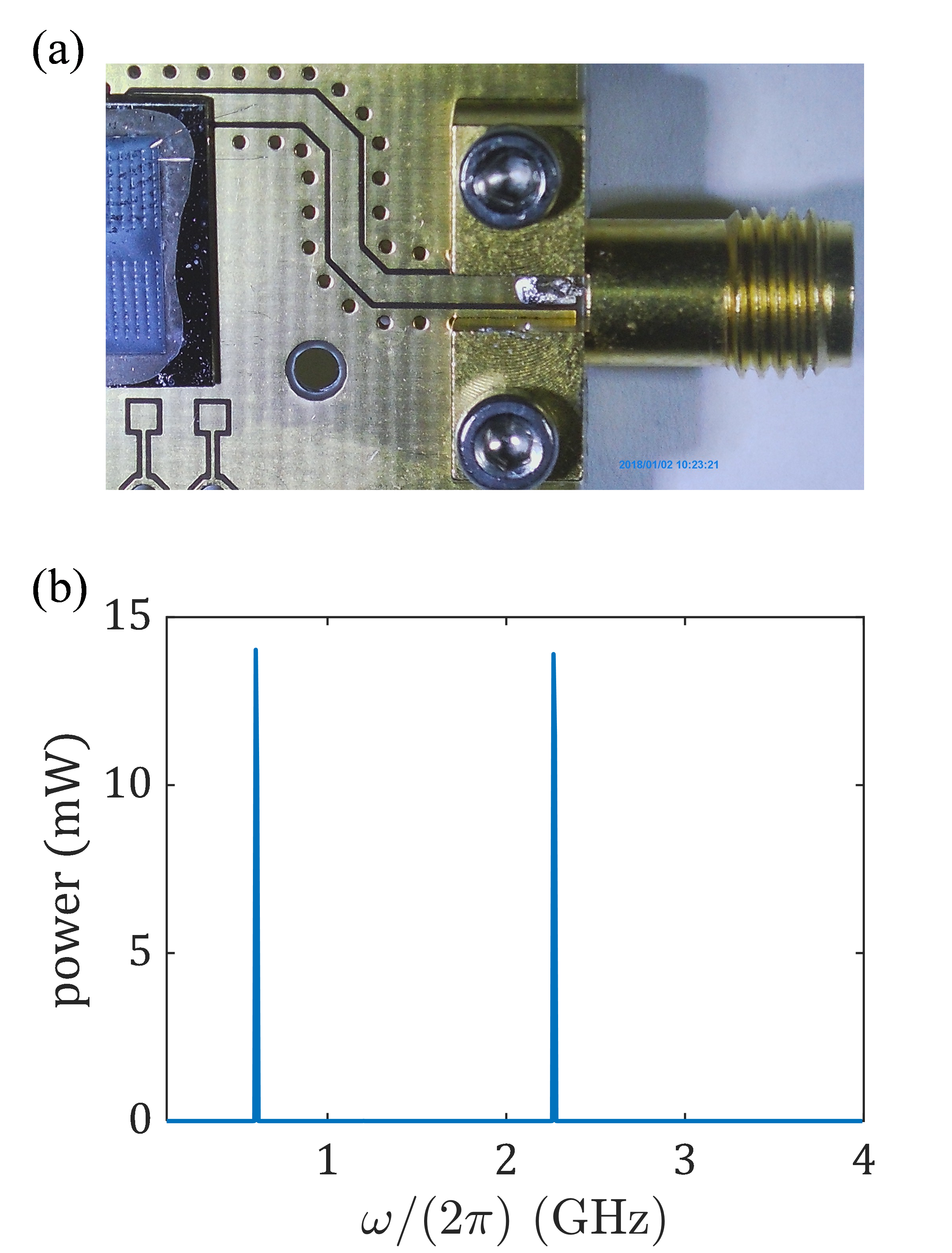}
\caption{(a) A photograph of the device wire bonded to a home-made PCB. The microwaves are fed into the waveguide via an SMA connector. (b) The spectrum of the two-frequency microwaves after passing through the power combiner, when $\omega_1 / (2\pi) = 0.6 \ \mathrm{GHz}$ and $\omega_2 / (2\pi) = 2.27 \ \mathrm{GHz}$. Both the output powers in the signal generators are set as $17 \ \mathrm{dBm}$.}
\label{fig6}
\end{figure}

\begin{figure}[t!]
\centering	\includegraphics[width=1\columnwidth]{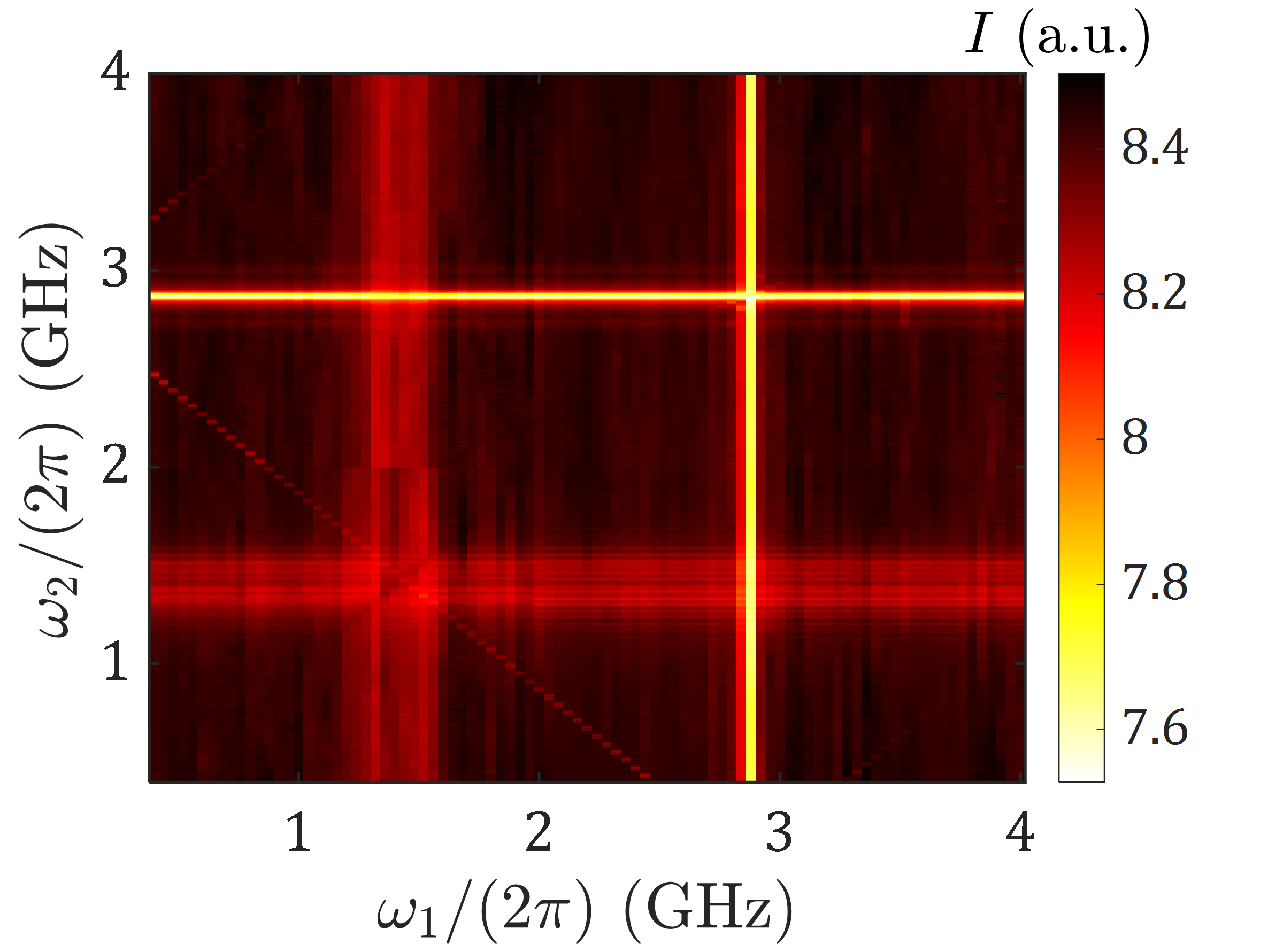}
\caption{The results of standard unmodulated continuous-wave ODMR measurements under driving microwaves of $\omega_1$ and $\omega_2$. The input microwave powers are $P_1 = P_2 = 13 \ \mathrm{mW}$.}
\label{fig7}
\end{figure}

\textit{Device fabrication}: For continuous-wave measurements, through standard photolithography followed by ion milling, we pattern a Cu(100 nm)/Pt(5 nm) stack on a silicon substrate into a straight microstrip with a length and width of 100 $\mathrm{\mu m}$ and 20 $\mathrm{\mu m}$, respectively, which is further wire bonded onto a home-made printed circuit board (PCB) with an SMA connector [Fig. \hyperref[fig6]{6(a)}]. Diamonds with a 1-$\mathrm{\mu m}$ diameter are dispersed on top of the microstrip. By calibrating the microwave signal, we determine that when the input power is 10 mW at the input terminal of the PCB, the microwave field is approximately 9 Oe at the microstrip surface. For the pulsed measurements on Rabi oscillations, a Cu(500 nm)/Pt(10 nm) stack on a sapphire substrate is patterned into a microstrip ring with an inner and outer diameter of 60 $\mathrm{\mu m}$ and 100 $\mathrm{\mu m}$, respectively, which is also wire bonded onto the same PCB. A single-crystal 15-$\mathrm{\mu m}$ diamond particle is placed at the center of the ring, to minimize the inhomogeneity of oscillating fields generated by microwaves. When the input microwave power is 200 mW, the field magnitude at the center of the ring is calibrated to be approximately 8 Oe.

\begin{figure*}[t!]
\centering	\includegraphics[width=1\textwidth]{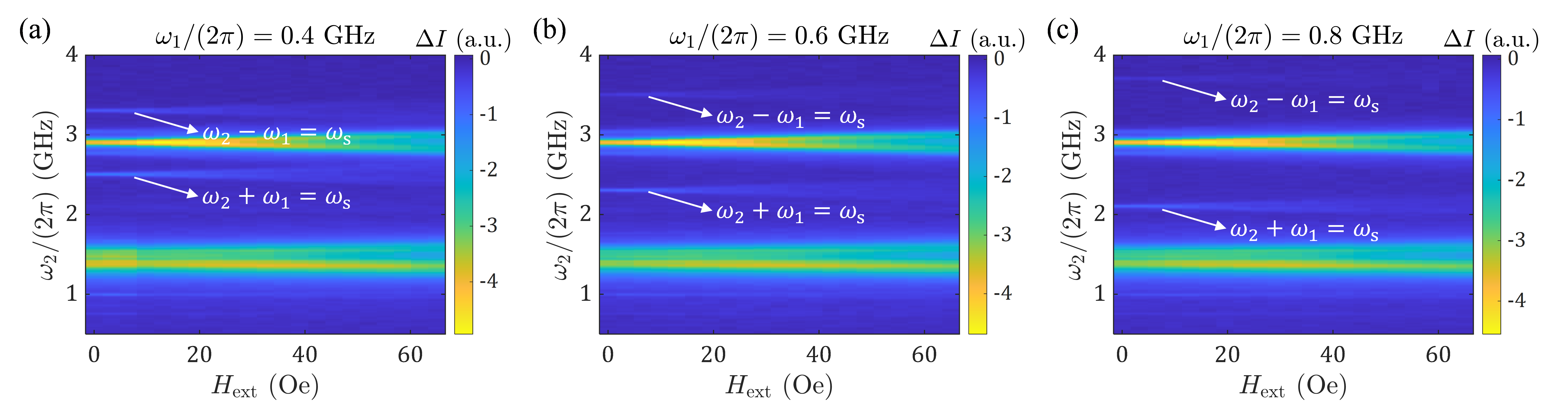}
\caption{$\Delta I$ as a function of $H_\mathrm{ext}$ when $\omega_2$ is swept and $\omega_1 / (2\pi)$ is fixed at (a) $0.4 \ \rm{GHz}$, (b) $0.6 \ \rm{GHz}$, and
(c) $0.8 \ \rm{GHz}$. Both the linear resonance dips at $\omega_2 = \omega_\mathrm{A, E}$ and the nonlinear dips at $\omega_2 \pm \omega_1 = \omega_\mathrm{A}$ manifest a cone-shaped structure with edges separated by $2 \gamma \mu_0 H_\mathrm{ext}$.}
\label{fig8}
\end{figure*}

\textit{Continuous-wave ODMR measurements}: The setup for continuous-wave ODMR measurements depicted in Fig. \hyperref[fig1]{1(d)} mainly consists of a home-built confocal microscope. To excite magnetic resonance in NV centers, continuous-wave signals are generated from two independent microwave signal generators (Anritsu 68369A and Anritsu 68347B) and then combined through a microwave power combiner (CentricRF CS6072). Green light from a $532$-$\rm nm$ DPSS laser is focused on the sample via a $1.25$-$\rm NA$ objective lens and illuminates NV centers in an ensemble of microdiamonds. The laser power is $0.6 \ \mathrm{mW}$ measured on the sample surface. We intentionally choose a low-power laser diode as the excitation light source, to enhance the ratio of resonance signals from high-order effects to those from linear effects by forcing the linear signals to approach saturation under limited optical pumping \cite{Jensen2013, Dreau2011}.The PL in the region of $600$-$800 \ \mathrm{nm}$ from NV centers is filtered and collected with a photomultiplier tube. To enhance the signal-to-noise ratio, the amplitude of one microwave source is modulated with a frequency of $104.42 \ \mathrm{Hz}$ and the change of the PL intensity is detected with a lock-in amplifier (EG\&G 7260). We have varied the position of the illuminated NV-center ensemble and found no qualitative difference in the ODMR results.

\textit{Linearity verification of the circuit}: With a spectrum analyzer (Anritsu MS8609A), we verify that the external microwave circuit acts purely linearly and is not the origin of frequency mixing. In Fig. \hyperref[fig6]{6(b)}, we show the spectrum of microwaves after they pass through the power combiner, when $\omega_1 / (2\pi) = 0.6 \ \mathrm{GHz}$ and $\omega_2 / (2\pi) = 2.27 \ \mathrm{GHz}$. Both the output powers in the signal generators are set as $17 \ \mathrm{dBm}$. Peaks only appear at $\omega_1$ and $\omega_2$, verifying the linearity of the external microwave circuit.

\textit{Comparison with unmodulated ODMR measurements}: In Fig. \ref{fig7}, we show the PL intensity $I$ under driving microwaves of $\omega_1$ and $\omega_2$, when we conduct a standard unmodulated continuous-wave measurement with an Agilent 34401A multimeter. Compared to Fig. \hyperref[fig2]{2(a)}, where the amplitude of the $\omega_2$ microwave is modulated with a low frequency, the unmodulated measurement has a much lower signal-to-noise ratio but we can still clearly identify the resonances at $\omega_2 + \omega_1 = \omega_\mathrm{A}$. The difference frequency resonances at $\pm(\omega_2 - \omega_1) = \omega_\mathrm{A}$ are buried in the noisy background to some degree but we can still distinguish them with extra attention. The feature of EIT, i.e., the dark line satisfying $\omega_1 = \omega_2$ when $\omega_{1,2}$ are both close to $\omega_\mathrm{E}$, is also observed in the unmodulated measurement. In conclusion, the unmodulated and modulated measurements show consistency in demonstrating the resonance at the sum or difference frequency, as well as the EIT effect. The modulated one gives a much higher signal-to-noise ratio and is therefore applied in the main experiment.

\textit{Field-dependent ODMR measurements}: The Zeeman splitting from a finite external static field $H_\mathrm{ext}$ (generated by an electromagnet in our experiments) will lift the degeneracy and result in distinct resonance frequencies for transitions from $\ket{m_s = 0}$ to $\ket{m_s = \pm1}$. In Figs. \hyperref[fig8]{8(a)–8(c)}, we show $\Delta I$ as a function of $H_\mathrm{ext}$ when $\omega_2$ is swept and $\omega_1 / (2\pi)$ is fixed at $0.4$, $0.6$, and $0.8 \ \rm{GHz}$, respectively. Due to random orientations of the NV-center principle axes with respect to the external field, the splitting in $\ket{m_s = \pm 1}$ manifests a cone-shaped structure in the resonance spectrum, as observed near $\omega_2 = \omega_\mathrm{A, E}$. In a similar vein, the nonlinear resonance near $\omega_2 \pm \omega_1 = \omega_\mathrm{A}$ also exhibits this conical feature, with edges separated by $2 \gamma \mu_0 H_\mathrm{ext}$.

\textit{Rabi-oscillation measurements}: The setup for two-photon Rabi-oscillation measurements is depicted in Fig. \hyperref[fig4]{4(a)}. We use an AWG (Feelelec FY8300) to modulate the laser and two microwave sources. The microwave sources and the power combiner are the same as the ones we use for continuous-wave measurements. The acousto-optic modulator that we use is the Isomet Model 1205C-1 with a driver of Model 532C. A permanent magnet is used to generate a static field of approximately 300 Oe at the position of the diamond sample, which helps us select one specific resonance frequency. The Rabi-oscillation data are obtained by recording the lock-in readout of the PL signals for varying pulse widths of microwaves.

\section{Solution of the density matrix for the NV spin under two driving microwaves}\label{appB}

When the external static field is zero, the spin sublevels $\ket{m_s = \pm1}$ in the NV optical ground state $\g$ or the excited state $\e$ are degenerate. In what follows, we derive the microwave photon induced spin transitions within $\g$ or $\e$. Throughout this section, we discuss the general case, in which $\omega_s$ represents the transition frequency between $\ket{m_s = \pm1}$ and $\ket{m_s = 0}$ either in the ground state $\g$ or in the excited state $\e$. Due to the symmetry between the $\ket{m_s = 0} \leftrightarrow \ket{m_s = +1}$, $\ket{m_s = 0} \leftrightarrow \ket{m_s = -1}$ transitions and their negligible mixing, we can focus on the $\ket{m_s = 0} \leftrightarrow \ket{m_s = +1}$ transition and consider the two-level spin system on the basis of $\ket{0} = \ket{m_s = 0}$ and $\ket{1} = \ket{m_s = +1}$. The intrinsic spin Hamiltonian is
\begin{equation}
    H_0 = \frac{\hbar \omega_s}{2} \sigma_z,
\label{H0}
\end{equation}

\noindent with reduced Planck constant $\hbar$ and Pauli matrices $\bm{\sigma} = (\sigma_x, \sigma_y, \sigma_z)$. We apply microwave fields
\begin{equation}
    \bm{h}(t) = \sum_{j=1,2} \bm{h}_j(t) = \sum_{j=1,2} h_j (\sin\theta \hat{x} + \cos\theta \hat{z}) \cos(\omega_j t + \varphi_j),
\label{h}
\end{equation}

\noindent where $\bm{h}_j(t)$ is the $j$th frequency component, $h_j$, $\omega_j$, and $\varphi_j$ are the amplitude, frequency, and phase of $\bm{h}_j(t)$, and $\theta$ is the angle between $\bm{h}_j(t)$ and the principal spin axis ($z$ axis) of the NV center ($j = 1, 2$). The interaction Hamiltonian between the spin and microwave photons is given by
\begin{equation}
    V(t) = \gamma\mu_0 \frac{\hbar}{2} \bm{\sigma} \cdot \bm{h}(t),
\label{V}
\end{equation}

\noindent with electron's gyromagnetic ratio $\gamma$ and vacuum permeability $\mu_0$. The total Hamiltonian is therefore given by
\begin{equation}
    H(t) = H_0 + V(t).
\label{H}
\end{equation}

\noindent The density matrix $\rho = \sum_{m,n=0,1} \rho_{mn}\ket{m}\bra{n}$ for the spin can be determined by solving the quantum master equation in the Lindblad form \cite{Jensen2013}:
\begin{equation}
    \frac{\partial \rho}{\partial t} = \frac{i}{\hbar} [\rho, H] + \sum_n \left(L_n \rho L_n^\dagger - \frac{1}{2} L_n^\dagger L_n \rho - \frac{1}{2} \rho L_n^\dagger L_n \right),
\label{Lindblad}
\end{equation}

\noindent where $L_n$ is the operator describing a nonunitary time evolution due to dissipative interactions between the spin and the environment. The longitudinal spin relaxation with a rate of $\Gamma_1^0$ and the transverse spin relaxation with a rate of $\Gamma_2^0$ can be described by
\begin{equation}
\begin{aligned}
    L_1 &= (\Gamma_1^0 / 2)^{1/2} \sigma_x, \\
    L_2 &= (\Gamma_2^0 / 2)^{1/2} \sigma_z.
\end{aligned}
\label{L1}
\end{equation}

\noindent Due to the photon bath of the laser pumping, $\ket{1}$ is optically pumped into $\ket{0}$ with a pump rate of $\Gamma_\mathrm{p}$, which can be described by
\begin{equation}
\begin{aligned}
    L_3 &= \Gamma_\mathrm{p}^{1/2} \ket{0} \bra{1}.
\end{aligned}
\label{L3}
\end{equation}

Combining Eqs. \eqref{H0}–\eqref{L3}, we obtain equations of motion for elements in the density matrix
\begin{small}
\begin{equation}
\begin{aligned}
    \frac{\partial {\rho}_{11}}{\partial t} &= - \left(\frac{\Gamma_1^0}{2} + \Gamma_\mathrm{p}\right) \rho_{11} + \frac{\Gamma_1^0}{2} \rho_{00} + \frac{i}{2} \gamma \mu_0 h_x (\rho_{01} - \rho_{10}), \\
    \frac{\partial {\rho}_{01}}{\partial t} &= (i \omega_s - \Gamma_2) \rho_{01} - i \gamma \mu_0 h_z \rho_{01} + \frac{i}{2} \gamma \mu_0 h_x (\rho_{11} - \rho_{00}),
\end{aligned}
\label{component}
\end{equation}
\end{small}

\noindent where $h_x = \sum_{j=1,2}h_j \sin\theta \cos(\omega_j t + \varphi_j)$ and $h_z = \sum_{j=1,2}h_j \cos\theta \cos(\omega_j t + \varphi_j)$ are the $x$-axis (transverse) and $z$-axis (longitudinal) components of $\bm{h}(t)$ and $\Gamma_2 = \Gamma_2^0 + \Gamma_\mathrm{p} / 2$ is the effective transverse spin relaxation rate. Other two elements in $\rho$ are determined by the constraints of $\rho_{10} = \rho_{01}^*$ and $\rho_{00} + \rho_{11} = 1$. We treat $V(t)$ as a perturbation to $H_0$ and solve for $\rho = \sum_{n=0}^{\infty} \rho^{(n)}$, where $\rho^{(n)}$ is in the $n$th order of $V(t)$ and is the $n$th-order correction to the zero-order solution $\rho^{(0)}$ given by $\rho_{11}^{(0)} = \Gamma_1^0 / (2 \Gamma_1)$ with the effective longitudinal spin relaxation rate $\Gamma_1 = \Gamma_1^0 + \Gamma_\mathrm{p}$ and $\rho_{01}^{(0)} = 0$. Specifically, we solve for $\rho^{(n)}$ ($n \geq 1$) using iterative equations
\begin{equation}
\begin{aligned}
    \frac{\partial {\rho}_{11}^{(n)}}{\partial t} = &- \Gamma_1 \rho_{11}^{(n)} + \frac{i}{2} \gamma \mu_0 h_x \left(\rho_{01}^{(n-1)} - \rho_{10}^{(n-1)}\right), \\
    \frac{\partial {\rho}_{01}^{(n)}}{\partial t} = \ &(i \omega_s - \Gamma_2) \rho_{01}^{(n)} - i \gamma \mu_0 h_z \rho_{01}^{(n-1)} + \\
    &\frac{i}{2} \gamma \mu_0 h_x \left(\rho_{11}^{(n-1)} - \rho_{00}^{(n-1)}\right).
\end{aligned}
\end{equation}

The first-order solution of $\rho$ is
\begin{equation}
\begin{aligned}
    \rho_{11}^{(1)} &= 0, \\
    \rho_{01}^{(1)} &= -\frac{\Gamma_\mathrm{p}\gamma \mu_0}{2\Gamma_1} \sum_{\omega_m} \frac{\Tilde{h}_x (\omega_m) e^{i \omega_m t}}{\omega_m - \omega_s - i\Gamma_2}.
\end{aligned}
\label{1st}
\end{equation}

\noindent Here, $\Tilde{h}_{x} (\omega)$ and $\Tilde{h}_{z} (\omega)$ are the Fourier transforms of $h_x$ and $h_z$, which are only finite at $\Tilde{h}_{x} (\pm \omega_j) = h_j \sin\theta e^{\pm i \varphi_j} / 2$ and $\Tilde{h}_{z} (\pm \omega_j) = h_j \cos\theta e^{\pm i \varphi_j} / 2$ where $j = 1, 2$. We define the first-order susceptibility $\chi_{xx}^{(1)}(\omega_j, \omega_j)$ through $\Tilde{M}_x(\omega_j) = \chi_{xx}^{(1)}(\omega_j, \omega_j) \Tilde{h}_x(\omega_j)$, where $M_x = \mathrm{Tr}(\rho \cdot \gamma\frac{\hbar}{2}\sigma_x) = \frac{1}{2}\gamma\hbar(\rho_{01} + \rho_{10})$ is the $x$-axis component of magnetic moment. It can be expressed as
\begin{equation}
\begin{aligned}
    \chi_{xx}^{(1)}(\omega_j, \omega_j) \approx -\frac{\Gamma_\mathrm{p} \gamma^2 \mu_0 \hbar}{4 \Gamma_1(\omega_j - \omega_s - i\Gamma_2)},
\end{aligned}
\label{chi1}
\end{equation}

\noindent when $|\omega_j - \omega_s| \ll |\omega_j + \omega_s|$ is satisfied ($j = 1, 2$).

The second-order solution of $\rho$ is
\begin{widetext}
\begin{equation}
\begin{aligned}
    \rho_{11}^{(2)} &= - \frac{\Gamma_\mathrm{p} \gamma^2 \mu_0^2}{4 \Gamma_1} \sum_{\omega_m, \omega_n} \frac{\Tilde{h}_x(\omega_m) \Tilde{h}_x(\omega_n) e^{i(\omega_m + \omega_n)t}}{(\omega_m + \omega_n - i\Gamma_1)(\omega_m - \omega_s - i\Gamma_2)} + \mathrm{H.c.}, \\
    \rho_{01}^{(2)} &= \frac{\Gamma_\mathrm{p} \gamma^2 \mu_0^2}{2 \Gamma_1} \sum_{\omega_m, \omega_n} \frac{\Tilde{h}_x(\omega_m) \Tilde{h}_z(\omega_n) e^{i(\omega_m + \omega_n)t}}{(\omega_m + \omega_n - \omega_s - i\Gamma_2)(\omega_m - \omega_s -i\Gamma_2)}.
\end{aligned}
\label{second}
\end{equation}
\end{widetext}

\noindent We define $\chi_{xxz}^{(2)}(\omega_2 \pm \omega_1, \omega_1, \omega_2)$ through $\Tilde{M}_x(\omega_2 \pm \omega_1) = \chi_{xxz}^{(2)}(\omega_2 \pm \omega_1, \omega_1, \omega_2) \Tilde{h}_x(\omega_1) \Tilde{h}_z(\omega_2)$ and obtain
\begin{widetext}
\begin{equation}
\begin{aligned}
    \chi_{xxz}^{(2)}(\omega_2 \pm \omega_1, \omega_1, \omega_2)
    \approx \ &\mp \frac{\Gamma_\mathrm{p} \gamma^3 \mu_0^2 \hbar \omega_s}{4 \Gamma_1 (\omega_2 \pm \omega_1 - \omega_s - i \Gamma_2) \omega_1 \omega_2},
\label{chi2}
\end{aligned}
\end{equation}
\end{widetext}

\noindent when $|\omega_2 \pm \omega_1 - \omega_s| \ll |\omega_2 \pm \omega_1 + \omega_s|$ and $\omega_{1, 2} \gg \Gamma_2$ are satisfied. The third-order solution of $\rho$ is
\begin{widetext}
\begin{equation}
\begin{aligned}
    \rho_{11}^{(3)} =
        \ & \frac{\Gamma_\mathrm{p} \gamma^3 \mu_0^3}{4\Gamma_1} \sum_{\omega_m, \omega_n, \omega_k}  
        \frac{\Tilde{h}_x(\omega_m) \Tilde{h}_z(\omega_n) \Tilde{h}_x(\omega_k) e^{i(\omega_m + \omega_n + \omega_k)t}}{(\omega_m + \omega_n - \omega_s - i \Gamma_2) (\omega_m - \omega_s - i \Gamma_2) (\omega_m + \omega_n + \omega_k - i \Gamma_1)} + \mathrm{H.c.}, \\
    \rho_{01}^{(3)} =
        -&\frac{\Gamma_\mathrm{p} \gamma^3 \mu_0^3}{4 \Gamma_1} \sum_{\omega_m, \omega_n, \omega_k}
        \frac{\Tilde{h}_x(\omega_m) \Tilde{h}_x(\omega_n) \Tilde{h}_x(\omega_k) e^{i(\omega_m + \omega_n + \omega_k)t}}{(\omega_m + \omega_n - i\Gamma_1) (\omega_m + \omega_n + \omega_k - \omega_s - i\Gamma_2)} \left(
        \frac{1}{\omega_m - \omega_s - i\Gamma_2} + \frac{1}{\omega_m + \omega_s - i\Gamma_2}
        \right) \\
        -&\frac{\Gamma_\mathrm{p} \gamma^3 \mu_0^3}{2 \Gamma_1} \sum_{\omega_m, \omega_n, \omega_k} \frac{\Tilde{h}_x(\omega_m) \Tilde{h}_z(\omega_n) \Tilde{h}_z(\omega_k) e^{i(\omega_m + \omega_n + \omega_k)t}}{(\omega_m + \omega_n - \omega_s - i\Gamma_2) (\omega_m - \omega_s - i\Gamma_2) (\omega_m + \omega_n + \omega_k - \omega_s - i\Gamma_2)}.
\end{aligned}
\label{3rd}
\end{equation}
\end{widetext}

\noindent We define $\chi_{xxxx}^{(3)}(\omega_2, \omega_2, -\omega_1, \omega_1)$ through $\Tilde{M}_x(\omega_2) = \chi_{xxxx}^{(3)}(\omega_2, \omega_2, -\omega_1, \omega_1) \Tilde{h}_x(\omega_2) \Tilde{h}_x(-\omega_1) \Tilde{h}_x(\omega_1)$ and obtain
\begin{widetext}
\begin{equation}
\begin{aligned}
    \chi_{xxxx}^{(3)}(\omega_2, \omega_2, -\omega_1, \omega_1) \approx \frac{i \gamma^4 \mu_0^3 \hbar \Gamma_\mathrm{p} \Gamma_2}{4 \Gamma_1 (\omega_2 - \omega_1 - i\Gamma_1) (\omega_2 - \omega_s - i\Gamma_2)^2 (-\omega_1 + \omega_s - i\Gamma_2)},
\end{aligned}
\label{chi3}
\end{equation}
\end{widetext}

\noindent when $|\omega_{1,2} - \omega_s| \ll |\omega_{1,2} + \omega_s|$ and $|\omega_1 - \omega_2| \ll \Gamma_2$ are satisfied. The fourth-order solution of $\rho_{11}^{(4)}$ is
\begin{widetext}
\begin{equation}
\begin{small}
\begin{aligned}
    \rho_{11}^{(4)} =
        -&\frac{\Gamma_\mathrm{p} \gamma^4 \mu_0^4}{8 \Gamma_1} \sum_{\omega_m, \omega_n, \omega_k, \omega_l} \left[\frac{\Tilde{h}_x(\omega_m) \Tilde{h}_x(\omega_n) \Tilde{h}_x(\omega_k) \Tilde{h}_x(\omega_l) e^{i(\omega_m + \omega_n + \omega_k + \omega_l)t}} {(\omega_m + \omega_n - i\Gamma_1) (\omega_m - \omega_s - i\Gamma_2) (\omega_m + \omega_n + \omega_k -\omega_s - i\Gamma_2) (\omega_m + \omega_n + \omega_k + \omega_l - i\Gamma_1)} + \mathrm{H.c.} \right] \\
        -&\frac{\Gamma_\mathrm{p} \gamma^4 \mu_0^4}{8 \Gamma_1} \sum_{\omega_m, \omega_n, \omega_k, \omega_l} \left[\frac{\Tilde{h}_x(\omega_m) \Tilde{h}_x(\omega_n) \Tilde{h}_x(\omega_k) \Tilde{h}_x(\omega_l) e^{i(\omega_m + \omega_n + \omega_k + \omega_l)t}} {(\omega_m + \omega_n - i\Gamma_1) (\omega_m + \omega_s - i\Gamma_2) (\omega_m + \omega_n + \omega_k - \omega_s - i\Gamma_2) (\omega_m + \omega_n + \omega_k + \omega_l - i\Gamma_1)} + \mathrm{H.c.} \right] \\
        -&\frac{\Gamma_\mathrm{p} \gamma^4 \mu_0^4}{4 \Gamma_1} \sum_{\omega_m, \omega_n, \omega_k, \omega_l} \left[\frac{\Tilde{h}_x(\omega_m) \Tilde{h}_z(\omega_n) \Tilde{h}_z(\omega_k) \Tilde{h}_x(\omega_l) e^{i(\omega_m + \omega_n + \omega_k + \omega_l)t}} {(\omega_m + \omega_n - \omega_s - i\Gamma_2) (\omega_m - \omega_s - i\Gamma_2) (\omega_m + \omega_n + \omega_k - \omega_s - i\Gamma_2) (\omega_m + \omega_n + \omega_k + \omega_l - i\Gamma_1)} + \mathrm{H.c.} \right].
\end{aligned}
\end{small}
\label{4th}
\end{equation}
\end{widetext}

\section{Origin of resonance at the sum or difference frequency of two driving microwaves}\label{appC}

The PL intensity is given by $I = I_0(1 - \alpha \braket{\rho_{11}})$, where $\braket{\rho_{11}}$ is the steady-state solution of $\rho_{11}$, $I_0$ is the PL intensity in the case of full spin initialization (i.e., $\braket{\rho_{11}} = 0$), and $\alpha$ is a phenomenological parameter to account for the difference in the contribution of the $\ket{m_s = 0}$ population and that of $\ket{m_s = \pm 1}$ population to PL intensity \cite{Dreau2011}. We note that the fast-oscillating terms of $\rho_{11}$ will not contribute to the PL intensity $I$.

For standard, linear magnetic resonance, the nonoscillating part in $\rho_{11}^{(2)}$ in Eq. \eqref{second} is given by
\begin{equation}
\begin{aligned}
    \braket{\rho_{11}^{(2)}} &\approx -\frac{\mu_0} {i \hbar \Gamma_1} \sum_{j=1,2} |\Tilde{h}_x(\omega_j)|^2 \chi_{xx}^{(1)}(\omega_j, \omega_j) + \mathrm{H.c.} \\ &= \frac{\gamma^2 \mu_0^2 \Gamma_\mathrm{p} \Gamma_2 \sin^2\theta}{8 \Gamma_1^2} \sum_{j=1,2} \frac{h_j^2}{(\omega_j - \omega_s)^2 + \Gamma_2^2},
\end{aligned}
\label{linearres}
\end{equation}

\noindent when $|\omega_{1, 2} - \omega_s| \ll |\omega_{1, 2} + \omega_s|$ are satisfied. Equation \eqref{linearres} demonstrates the commonly observed resonance under a single microwave frequency $\omega_1$ or $\omega_2$ close to $\omega_s$.

Next we consider nonlinear, two-photon magnetic resonance. The third term on the right-hand side of Eq. \eqref{4th} gives rise to the observed resonance when the sum or difference frequency of $\omega_1$ and $\omega_2$ matches $\omega_\mathrm{A}$. When $\omega_{1,2}$ are individually far away from $\omega_\mathrm{A}$ but their sum or difference is around the resonance condition of $\omega_2 \pm \omega_1 = \omega_\mathrm{A}$, we have
\begin{equation}
    \braket{\rho_{11}^{(4)}} \approx \frac{|\chi_{xxz}^{(2)}(\omega_2 \pm \omega_1, \omega_1, \omega_2)|^2 h_1^2 h_2^2 \sin^2(2\theta) \Gamma_\mathrm{2, A}}{8 \gamma^2 \hbar^2 \Gamma_\mathrm{p}},
\end{equation}

\noindent where $\chi_{xxz}^{(2)}$ is in the expression of Eq. \eqref{chi2}, with $\omega_s$, $\Gamma_1$, and $\Gamma_2$ replaced by $\omega_\mathrm{A}$, $\Gamma_\mathrm{1, A}$, and $\Gamma_\mathrm{2, A}$, and we have utilized the approximations of $\omega_2 \pm \omega_1 \approx \omega_\mathrm{A}$ and $\omega_{1, 2} \gg \Gamma_\mathrm{2, A}$. After averaging $\theta$ over $[0, \pi]$ (note that the detected NV centers in the ensemble have randomly oriented principal axes), the change of the PL intensity compared to that without microwaves, defined as $\Delta I = I_0 - I_0(1 - \alpha \braket{\rho_{11}^{(0)}})$, can be expressed as
\begin{equation}
\begin{aligned}
    \Delta I &\approx -|\Delta I|_\mathrm{max} \braket{\rho_{11}^{(4)}} \\
    &= - \eta |\chi_{xxz}^{(2)}(\omega_2 \pm \omega_1, \omega_1, \omega_2)|^2 h_1^2 h_2^2,
\end{aligned}
\label{Delta I}
\end{equation}

\noindent where $|\Delta I|_\mathrm{max} = I_0 \alpha$ is the difference of the PL intensity between the case of full spin initialization with $\braket{\rho_{11}} = 0$ and the case of full spin inversion with $\braket{\rho_{11}} = 1$ ($|\Delta I|_\mathrm{max} = I_0 \alpha \epsilon$ with modulation depth $\epsilon$ if we consider amplitude-modulated measurements) and
\begin{equation}
    \eta = \frac{|\Delta I|_\mathrm{max} \Gamma_\mathrm{2, A}}{16 \gamma^2 \hbar^2 \Gamma_\mathrm{p}}
\label{eta}
\end{equation}

\noindent is a constant factor. We see that $\Delta I$ in Eq. \eqref{Delta I} is a Lorentzian with a line width of $\Gamma_\mathrm{2, A}$ and a contrast of
\begin{equation}
    C = \frac{|\Delta I|_\mathrm{max} \Gamma_\mathrm{p} \omega_\mathrm{A}^2 \gamma^4 \mu_0^4 h_1^2 h_2^2}{256 \Gamma_\mathrm{1, A}^2 \Gamma_\mathrm{2, A} \omega_1^2 \omega_2^2}.
\label{peak height}
\end{equation}

\noindent Since $\Gamma_\mathrm{2, E}$ is much larger than $\Gamma_\mathrm{2, A}$, as verified by the broad resonance dip at $\omega_2 = \omega_\mathrm{E}$ [see Figs. \hyperref[fig2]{2(a)} and \ref{fig7}], the resonance at the sum or difference frequency of $\omega_{1,2}$ is significant in $\g$ but not observed in $\e$.

Up to now, the line width of the sum- or difference-frequency resonance, i.e., the effective transverse spin relaxation rate, has been determined to be $\Gamma_\mathrm{2, A} = \Gamma_\mathrm{2, A}^0 + \Gamma_\mathrm{p} / 2$ for a single NV center or an ensemble of totally identical NV centers. In reality, however, NV centers in the ensemble are different, since they are situated in varying local environments, which leads to inhomogeneous broadening. We can model this by assuming that the transition frequency $\omega_\mathrm{A}$ is distributed based on the probability distribution function $p(\omega_\mathrm{A})$. Then, the detected change of the PL intensity averaged on the ensemble is $\langle \Delta I \rangle = \int_0^\infty \Delta I \cdot p(\omega_\mathrm{A}) d\omega_\mathrm{A}$, where $\Delta I$ is given by Eq. \eqref{Delta I}. Typically, it is assumed that $p(\omega_\mathrm{A})$ is Gaussian \cite{Dobrovitski2008} and $\langle \Delta I \rangle$ is in the Voigt line shape, the line width of which can be calculated by some complex error functions. Here, we simply assume $p(\omega_\mathrm{A})$ is a Lorentzian, with a line width of $\Gamma_\mathrm{inh}$. In this case, $\langle \Delta I \rangle$ is also a Lorentzian, with a total line width of $\Gamma_\mathrm{2, A} = \Gamma_\mathrm{2, A}^0 + \Gamma_\mathrm{p} / 2 + \Gamma_\mathrm{inh}$. Since $\Delta I$ around the resonance condition in our experiment is almost a perfect Lorentzian [see Fig. \hyperref[fig3]{3(a)}], this assumption is reasonable.  

\section{Analysis on higher-order resonance dips}\label{appD}

In Fig. \hyperref[fig2]{2(a)}, we observe additional bright lines (labeled from 4 to 7), which can be traced to magnetic resonance excited by even higher-order processes. Here, to examine whether our theory generally applies to higher-order resonance dips, we select the three-photon resonance at $2\omega_2 - \omega_1 = \omega_\mathrm{A}$ for a power- and frequency-dependent test. From Eq. \eqref{3rd}, the third-order susceptibility $\chi_{xxxx}^{(3)}(2\omega_2 - \omega_1, \omega_2, \omega_2, -\omega_1)$, defined through

\begin{widetext}
\begin{equation}
    \Tilde{M}_x(2\omega_2 - \omega_1) = \chi_{xxxx}^{(3)}(2\omega_2 - \omega_1, \omega_2, \omega_2, -\omega_1) \Tilde{h}_x(\omega_2) \Tilde{h}_x(\omega_2) \Tilde{h}_x(-\omega_1)    
\end{equation}
\end{widetext}

\noindent and $\chi_{xxzz}^{(3)}(2\omega_2 - \omega_1, \omega_2, \omega_2, -\omega_1)$, defined through

\begin{widetext}
\begin{equation}
    \Tilde{M}_x(2\omega_2 - \omega_1) = \chi_{xxzz}^{(3)}(2\omega_2 - \omega_1, \omega_2, \omega_2, -\omega_1) \Tilde{h}_x(\omega_2) \Tilde{h}_z(\omega_2) \Tilde{h}_z(-\omega_1)  
\end{equation}
\end{widetext}

\noindent are given by
\begin{widetext}
\begin{equation}
\begin{aligned}
    \chi_{xxxx}^{(3)}(2\omega_2 - \omega_1, \omega_2, \omega_2, -\omega_1) &\approx \frac{i \gamma^4 \mu_0^3 \hbar \Gamma_\mathrm{p} \Gamma_2 \omega_s}{8 \Gamma_1 \omega_2(\omega_s - \omega_2)(\omega_s - \omega_1) (2\omega_2 - \omega_1 - \omega_s - i\Gamma_2)}, \\
    \chi_{xxzz}^{(3)}(2\omega_2 - \omega_1, \omega_2, \omega_2, -\omega_1) &\approx \frac{i \gamma^4 \mu_0^3 \hbar \Gamma_\mathrm{p} \Gamma_2}{4 \Gamma_1 \omega_1 \omega_2 (2\omega_2 - \omega_1 - \omega_s - i\Gamma_2)},
\end{aligned}
\label{chi3-1}
\end{equation}
\end{widetext}

\noindent respectively, when the near-resonance condition of $2\omega_2 - \omega_1 \approx \omega_s$ is satisfied. It is noted that $\chi_{xxxx}^{(3)}$ represents the spin transition from $\ket{m_s = 0}$ to $\ket{m_s = \pm1}$ via $\sigma_\pm + \sigma_\pm - \sigma_\pm$. The transition rate is proportional to $(h_{2x}^2 h_{1x} \omega_s / [\omega_2(\omega_s - \omega_2)(\omega_s - \omega_1)]$, where $h_{jx} = h_j \cos\theta_j$ and $h_{jz} = h_j \sin\theta_j$ ($j = 1, 2$). On the other hand, $\chi_{xxzz}^{(3)}$ represents the spin transition via $\sigma^\pm + \pi + \pi$. The transition rate is proportional to $h_{2x} h_{2z} h_{1x} / (\omega_1 \omega_2)$. For a general case with $\theta \neq 0, \pi/2, \pi$, both of the two kinds of three-photon spin transitions will contribute to the detected resonance signals. Note that in the visited frequency ranges in Fig. \hyperref[fig3]{3(h)}, both of $\omega_{1, 2}$ are close to $\omega_\mathrm{A}$. Consequently, $\omega_\mathrm{A} / [\omega_2(\omega_\mathrm{A} - \omega_2)(\omega_\mathrm{A} - \omega_1)]$, the frequency dependence from $\chi_{xxxx}^{(3)}$, is more significant than $1 / (\omega_1 \omega_2)$, the frequency dependence from $\chi_{xxzz}^{(3)}$, which is demonstrated by dashed lines in Fig. \hyperref[fig3]{3(h)}. The semiquantitative consistency between the experimental data on the third-order resonance and our theoretical model suggests the generality of the model to higher-order resonances.

\begin{figure*}[t!]
\centering	\includegraphics[width=1\textwidth]{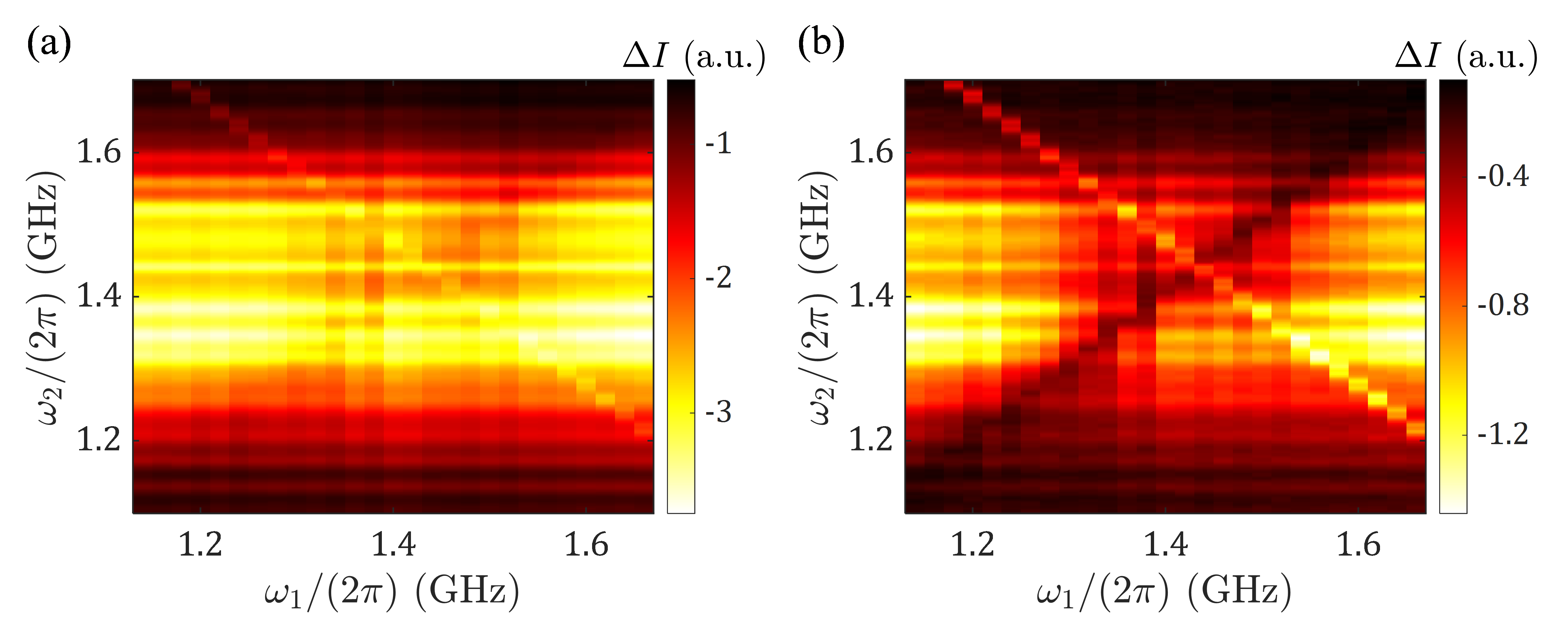}
\caption{$\Delta I$ under $\omega_1$ and $\omega_2$ inputs when $\omega_{1,2}$ are close to $\omega_\mathrm{E}$: (a) microwave powers $P_1 = 5 \ \mathrm{mW}$, $P_2 = 13 \ \mathrm{mW}$; (b) microwave powers $P_1 = 13 \ \mathrm{mW}$, $P_2 = 5 \ \mathrm{mW}$. Only (b) shows the dark line satisfying $\omega_1 = \omega_2$, which is a feature of EIT.}
\label{fig9}
\end{figure*}

\section{Origin of electromagnetically induced transparency}\label{appE}

In Figs. \hyperref[fig2]{2(a)} and \hyperref[fig7]{7}, on top of the series of the bright resonance lines, we observe a dark line satisfying $\omega_1 = \omega_2$ within the broad resonance dip at $\omega_2 = \omega_\mathrm{E}$. This feature of magnetic resonance suppression under zero detuning of two waves is very similar to the EIT phenomenon studied in nonlinear optics \cite{Fleischhauer2005}, where in the presence of a strong pump wave, the interaction between the probe wave and the matter is minimized due to interference effects. Mathematically, the first two terms on the right-hand side of Eq. \eqref{4th} play an important role. Combining Eqs. \eqref{4th} and \eqref{linearres}, when $\omega_1 \approx \omega_2$ and both of them are close to $ \omega_\mathrm{E}$, i.e., $|\omega_{1,2} - \omega_\mathrm{E}| \ll |\omega_{1,2} + \omega_\mathrm{E}|$ and $|\omega_1 - \omega_2| \ll \Gamma_\mathrm{2, E}$ are satisfied, we have
\begin{widetext}
\begin{equation}
\begin{aligned}
    \braket{\rho_{11}^{(2)}} + \braket{\rho_{11}^{(4)}} \approx -\frac{\mu_0 |\tilde{h}_x(\omega_2)|^2}{i \Gamma_\mathrm{1, E} \hbar} \left[
    \chi_{xx}^{(1)}(\omega_2, \omega_2) + |\Tilde{h}_x(\omega_1)|^2 \chi_{xxxx}^{(3)}(\omega_2, \omega_2, -\omega_1, \omega_1)
    \right] + \mathrm{H.c.} + (\omega_1 \Longleftrightarrow \omega_2).
\end{aligned}
\end{equation}
\end{widetext}

\noindent where $\chi_{xx}^{(1)}$ and $\chi_{xxxx}^{(3)}$ are in the expressions of Eqs. \eqref{chi1} and \eqref{chi3}, with $\omega_s$, $\Gamma_1$, and $\Gamma_2$ replaced by $\omega_\mathrm{E}$, $\Gamma_\mathrm{1, E}$, and $\Gamma_\mathrm{2, E}$. After averaging $\theta$ over $[0, \pi]$ and only keeping terms related to $h_2$, since only the amplitude of the $\omega_2$ input is modulated, the destructive interference between the spin transitions associated with $\chi_{xx}^{(1)}$ and $\chi_{xxxx}^{(3)}$ results in
\begin{equation}
\begin{aligned}
    \Delta I &\approx -|\Delta I|_\mathrm{max} \left[\braket{\rho_{11}^{(2)}} + \braket{\rho_{11}^{(4)}}\right] \\ &=
    \frac{iA}{\omega_2 - \omega_\mathrm{E} - i\Gamma_\mathrm{2, E} - \frac{B}{\omega_2 - \omega_1 - i\Gamma_\mathrm{1, E}}} + \mathrm{H.c.}
\label{EIT}
\end{aligned}
\end{equation}

\noindent where $A = |\Delta I|_\mathrm{max} \Gamma_\mathrm{p} \gamma^2 \mu_0^2 h_2^2 / (32\Gamma_\mathrm{1, E}^2)$ and $B = i\Gamma_\mathrm{2,E}\gamma^2\mu_0^2 h_1^2 / [4 (\omega_1 - \omega_\mathrm{E} + i\Gamma_\mathrm{2,E})] \approx \gamma^2\mu_0^2 h_1^2 / 4$. We see that the peak value of $|\Delta I|$ is suppressed at zero detuning $\omega_2 - \omega_1 = 0$, and that the EIT effect is most significant when $\Gamma_\mathrm{2, E} > \gamma \mu_0 h_1 / 2 > \Gamma_\mathrm{1, E}$, which is satisfied in the spin transitions in $\e$. Comparatively, the relatively smaller $\Gamma_\mathrm{2,A}$ makes the EIT feature in $\g$ less noticeable. We note that here, we treat the $\omega_1$ signal as the probe wave and the modulated $\omega_2$ signal as the pump wave. When using a weak pump power and a high probe power [$P_1 = 5 \ \mathrm{mW}$, $P_2 = 13 \ \mathrm{mW}$; Fig. \hyperref[fig9]{9(a)}], the transparency window is not observed. In contrast, when using a high pump power and a weak probe power [$P_1 = 13 \ \mathrm{mW}$, $P_2 = 5 \ \mathrm{mW}$; Fig. \hyperref[fig9]{9(b)}], the transparency window clearly exists. This asymmetric result is consistent with Eq. \eqref{EIT}, further verifying the EIT origin of the resonance suppression feature.

The derivation from Appendixes \ref{appB}–\ref{appE} applies to the two-level model for the transition between $\ket{m_s = 0}$ and $\ket{m_s = +1}$. When we focus on the transition between $\ket{m_s = 0}$ and $\ket{m_s = -1}$, the treatment is similar. The only difference in the derivation is to replace $h_z$ by $-h_z$ due to the negative magnetic moment, which will add an extra negative sign in the results of density-matrix elements that are odd functions of $h_z$ (e.g., $\rho_{01}^{(2)}$ and $\rho_{11}^{(3)}$). The elements $\rho_{11}^{(2)}$, $\rho_{11}^{(4)}$, and hence $\Delta I$ will not be affected.

\begin{figure*}[t!]
\centering	\includegraphics[width=1\textwidth]{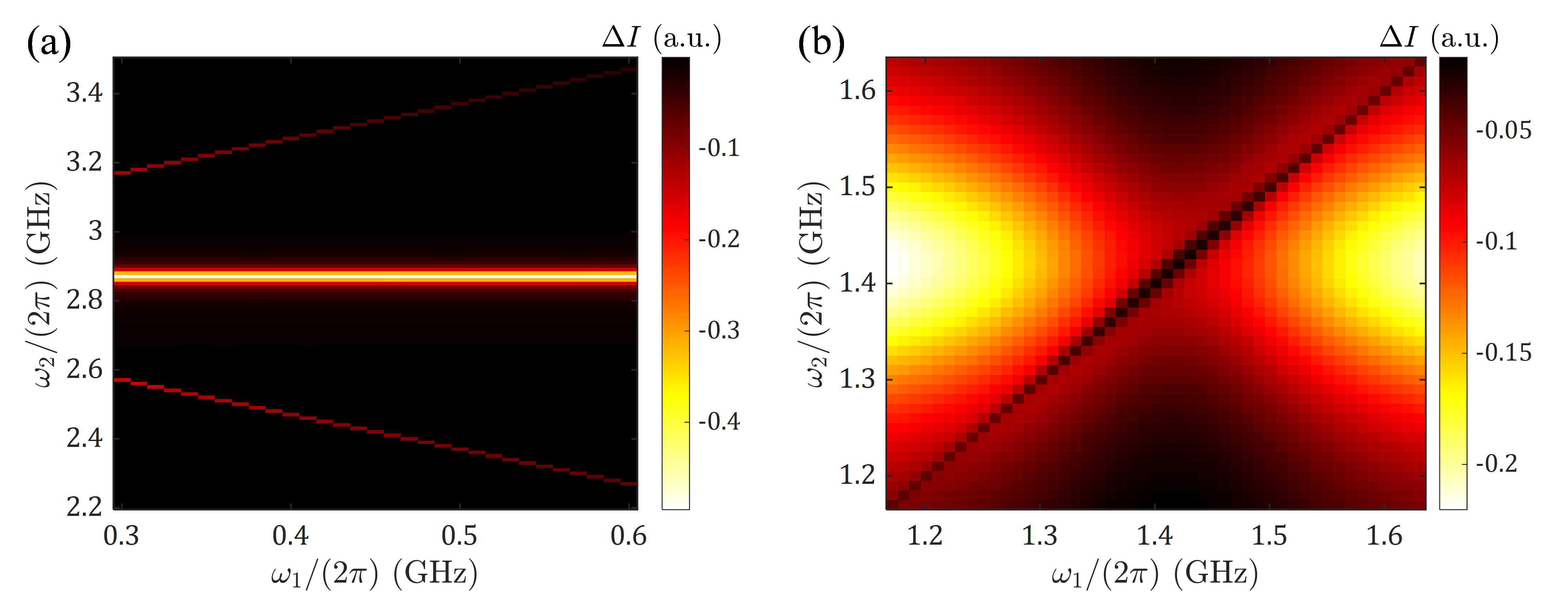}
\caption{Numerically simulated results demonstrating (a) the sum- and difference-frequency resonances at $\omega_2 \pm \omega_1 = \omega_\mathrm{A}$ and (b) EIT effect when $\omega_{1,2}$ are close to $\omega_\mathrm{E}$.}
\label{fig10}
\end{figure*}

\section{Numerical simulations on the density-matrix master equation}\label{appF}

Besides the theoretical derivation based on perturbation theory, we can also solve Eq. \eqref{component} numerically. In Figs. \hyperref[fig10]{10(a)} and \hyperref[fig10]{10(b)}, we show the numerically simulated results for the sum- and difference-frequency resonances at $\omega_2 \pm \omega_1 = \omega_\mathrm{A}$ and the EIT effect when $\omega_{1,2}$ are close to $\omega_\mathrm{E}$, respectively. The parameters used in the simulations are listed below.

Figure \hyperref[fig10]{10(a)}: $\omega_\mathrm{A} / (2\pi) = 2.87 \ \mathrm{GHz}$, $\Gamma_\mathrm{1,A}^0 / (2\pi) = 1 \ \mathrm{kHz}$, $\Gamma_\mathrm{2,A}^0 / (2\pi) = 5 \ \mathrm{MHz}$, $\Gamma_\mathrm{p} / (2\pi) = 3 \ \mathrm{MHz}$, $h_{1, 2} = 20 \ \mathrm{Oe}$.

Figure \hyperref[fig10]{10(b)}: $\omega_\mathrm{E} / (2\pi) = 1.42 \ \mathrm{GHz}$, $\Gamma_\mathrm{1,E}^0 / (2\pi) = 1 \ \mathrm{kHz}$, $\Gamma_\mathrm{2,E}^0 / (2\pi) = 100 \ \mathrm{MHz}$, $\Gamma_\mathrm{p} / (2\pi) = 3 \ \mathrm{MHz}$, $h_{1,2} = 20 \ \mathrm{Oe}$.

Here, $\Gamma_\mathrm{2,A}^0$ and $\Gamma_\mathrm{2,E}^0$ include the contribution of ensemble inhomogeneous broadening. $\theta$ is set as $45^\circ$ to reflect the effect of the ensemble average. $|\Delta I|_\mathrm{max}$ is set as 1. We take a time step of $\Delta t = 2 \times 10^{-13} \ \mathrm{s}$ and average $\rho_{11}(t)$ in $4 \times 10^{-7} \ \mathrm{s} \leq t \leq 8 \times 10^{-7} \ \mathrm{s}$ to obtain steady-state solutions. The results with the same parameters except $h_2 = 0$ are subtracted to imitate the modulation on $h_2$. In Figs. \hyperref[fig10]{10(a)} and \hyperref[fig10]{10(b)}, we see that the numerical simulations are consistent with the experimental data in clearly demonstrating features induced by nonlinear multiphoton processes.

\bibliography{main_text}

\end{document}